\documentclass[%
reprint,
 amsmath,amssymb,
 aps,
 30pt,
]{revtex4-1}

\usepackage{graphicx}% Include figure files
\usepackage[caption=false]{subfig}
\usepackage{dcolumn}% Align table columns on decimal point
\usepackage{bm}% bold math
\usepackage{xcolor}
\usepackage{soul}
\usepackage[colorlinks=true,linkcolor=blue,filecolor=blue,menucolor=yellow,urlcolor=blue,
citecolor=blue,anchorcolor=blue]{hyperref}
\usepackage[export]{adjustbox}
\usepackage{floatrow}

\usepackage{setspace}

\begin{document}

%\renewcommand{\thefootnote}{\fnsymbol{footnote}}

%\title{Dual Role of Cell-Cell Adhesion In Tumor Suppression and Proliferation}
%
%\author{Abdul N Malmi-Kakkada$^{a}$, Xin Li$^{a}$, Sumit Sinha$^{b}$, D. Thirumalai$^{a}$
%\footnote{Corresponding author: dave.thirumalai@gmail.com}}
%\affiliation{$^{a}$Department of Chemistry, University of Texas at Austin, Austin, TX 78712, \\ $^{b}$Department of Physics, University of Texas at Austin, Austin, TX 78712}

\title{Dual Role of Cell-Cell Adhesion In Tumor Suppression and Proliferation Due to Collective Mechanosensing}

% Use letters for affiliations, numbers to show equal authorship (if applicable) and to indicate the corresponding author
\author{Abdul N Malmi-Kakkada$^1$, Xin Li$^1$, Sumit Sinha$^2$, D. Thirumalai$^1$}
\email{dave.thirumalai@gmail.com}
\affiliation{$^1$Department of Chemistry, University of Texas, Austin, TX 78712, USA.}
\affiliation{$^2$Department of Physics, University of Texas, Austin, TX 78712, USA.}

%\vspace{2cm}
\date{\today}

\begin{abstract}
{\bf Abstract:} It is known that 
mechanical interactions  couple a cell to its neighbors, enabling a feedback loop  
to regulate tissue growth. However, the interplay between cell-cell adhesion strength, local cell density 
and force fluctuations in regulating cell proliferation is poorly understood.  Here, we show that spatial variations in the tumor growth rates, 
which depend on the location of cells within tissue spheroids, 
are strongly influenced by cell-cell adhesion. As the strength of the cell-cell adhesion increases, 
intercellular pressure initially decreases, enabling dormant cells 
to more readily enter into a proliferative state. We identify an optimal cell-cell adhesion regime  
where pressure on a cell is a minimum, allowing for maximum 
proliferation. We use a theoretical model to validate this novel collective feedback mechanism coupling adhesion strength,  
local stress fluctuations and proliferation.
Our results predict the existence of a non-monotonic proliferation behavior as 
a function of adhesion strength, consistent with experimental 
results. Several experimental implications of the proposed role of cell-cell adhesion 
in proliferation are quantified, making our model predictions amenable to further experimental scrutiny. 
We show that the mechanism of contact inhibition of proliferation, based on 
a pressure-adhesion feedback loop, serves as a unifying mechanism to understand 
the role of cell-cell adhesion in proliferation. 
\end{abstract}
%\newpage
\maketitle

\section{Introduction}
Adhesive forces between cells, mediated by cadherins, play a critical role in 
morphogenesis, tissue healing, and tumor growth~\cite{legoff2016mechanical,friedl2017tuning}. 
In these processes, the collective cell properties are influenced by how cells adhere to one another, 
enabling cells to communicate through mechanical  
forces~\cite{budnar2013mechanobiological,tambe2011collective}. 
Amongst the family of cadherins, E-cadherin is the most abundant, found in 
most metazoan cohesive tissues~\cite{halbleib2006cadherins}. E-cadherin transmembrane proteins 
facilitate intercellular bonds through the interaction of extracellular domains on neighboring cells. 
The function of cadherins was originally appreciated through their role in cell aggregation  
during morphogenesis~\cite{takeichi1988cadherins,saito2012classical}. 
Mechanical coupling between the cortical cytoskeleton and cell membrane is understood to involve  
the cadherin cytoplasmic domain~\cite{tabdanov2009role}. Forces exerted across 
cell-cell contacts is transduced between cadherin extracellular domain and the cellular cytoskeletal 
machinery through the cadherin/catenin complex~\cite{borghi2012cadherin}. Therefore, to 
understand how mechanical forces control the spatial organization of cells within 
tissues and impact proliferation, the role of adhesion strength at cell-cell contacts need to be elucidated.  

Together with cell-cell adhesion, cell proliferation control is of 
fundamental importance in animal and plant development, regeneration, and 
cancer progression~\cite{thompson1942growth,shaw2009wound}.
Spatial constraints due to cell packing or crowding are known to affect cell 
proliferation~\cite{abercrombie1970contact,folkman1978role,chen1997geometric,shraiman2005mechanical,streichan2014spatial,jacobeen2018cellular}.
The spatiotemporal arrangement of cells in response to local 
stress field fluctuations, arising from intercellular interactions, 
and how it feeds back onto cell proliferation % in turn arising from local density fluctuations and how it feeds ... 
remains unclear. Indeed, evidence so far based on experimental and theoretical studies 
on the mechanism underlying the cross-talk between the strength of cell-cell adhesion and  
proliferation, invasion and drug resistance is not well understood~\cite{shraiman2005mechanical,puliafito2012collective,mcclatchey2012contact,kourtidis2017central,irvine2017mechanical}. 

We briefly discuss two seemingly paradoxical roles of E-cadherin in proliferation. 
E-cadherin depletion lead cells to adopt a mesenchymal morphology characterized by 
enhanced cell migration and invasion~\cite{vleminckx1991genetic,perl1998causal} while 
increased expression of E-cadherin in cell lines with minimal expression reverses  
highly proliferative and invasive phenotypes~\cite{frixen1991cadherin,wong2003adhesion}. 
Besides %being a proliferation suppressor there is also evidence that cadherin expression can inhibit cell growth rate~\cite{watabe1994induction,croix1998cadherin}. 
suppressing tumor growth, there is also evidence that cadherin expression can lead to tumor proliferation. 
We detail the dual role that E-cadherin plays below. 

{\it E-cadherin downregulation and tumor progression through the EMT mechanism:} Loss of E-cadherin expression is related to 
the epithelial-to-mesenchymal transition (EMT), observed during embryogenesis~\cite{takeichi1988cadherins,gumbiner1996cell}, 
tumor progression and metastasis~\cite{cano2000transcription,yang2008epithelial}. 
EMT results in the transformation of 
epithelial cells into a mesenchymal phenotype, where adhesive strength between the cells is significantly decreased,   %and upregulated expression of other types of 
and drives tumor invasiveness and 
cell migration in epithelial tumors. In this `canonical' picture, down regulation of cell-cell adhesion contributes to 
cancer progression~\cite{weinberg2013biology}.  

{\it E-cadherin upregulation may promote tumor progression:} %Even though E-cadherin loss is considered to 
%be a critical driver in cancer progression, 
In contrast to the reports cited above, others argue 
that EMT may not be required for %either tumorigenesis or 
cancer metastasis~\cite{lou2008epithelial,fischer2015emt,zheng2015emt}. 
In fact, E-cadherin may facilitate collective cell migration 
that potentiates invasion and metastasis~\cite{cheung2013collective,shamir2014twist1}. 
%In terms of aiding cell proliferation, 
Increased E-cadherin expression is deemed necessary for the progression of aggressive tumors such 
as inflammatory breast cancer (IBC)~\cite{silvera2009essential} 
and a glioblastoma multiforme (GBM) subtype~\cite{lewis2010misregulated,rodriguez2012cadherin}, and invasive ductal carcinomas~\cite{padmanaban2019cadherin}. 
In multiple GBM cell lines, increased E-cadherin expression positively correlated with tumor growth and invasiveness~\cite{lewis2010misregulated}. 
In normal rat kidney-52E (NRK-52E) and non-tumorigenic human mammary epithelial cells (MCF-10A), %change1
E-cadherin engagement stimulated a peak in proliferation capacity through Rac1 activation~\cite{liu2006cadherin}. 
By culturing cells in micro-fabricated wells to control for cell spreading on 2D substrates, 
VE-cadherin mediated contact with neighboring cells showed enhanced proliferation~\cite{gray2008engineering}. 
The dual  role of cell-cell adhesion, suppressing proliferation in some cases and promoting
it in other instances, therefore warrants further investigation. 

The contrasting scenario raise unanswered questions that are amenable to analyses using relatively simple models of tumor growth:  
(1) How does the magnitude of forces exerted by cells on one another influence their overall growth and proliferation? 
(2) Can a minimal physical model capture the role of cell-cell adhesion in suppressing and enhancing tumor growth?  
Here we use simulations and theoretical arguments to establish a 
non-monotonic dependence between cell-cell adhesion strength 
($f^{ad}$; depth of attractive interaction between cells) and proliferation capacity. 
%, allowing us %change2
%to explain the apparently contradictory results detailed above. 
While the parameter $f^{ad}$ is a proxy for  
cell-cell adhesion strength representing E-cadherin expression, we note that adhesion strength may also  
increase due to cadherin clustering~\cite{yap2015adherens}, increasing time of contact between cells~\cite{yap2015adherens}, or through ``mechanical 
polarization'', where cells reorganize their adhesive and cytoskeletal machinery to suppress actin density 
along cell-cell contact interfaces~\cite{amack2012knowing,david2014tissue}.
We show that cell proliferation increases as $f^{ad}$ increases, % a proxy for E-cadherin expression, 
and reaches a maximum at a critical value, $f^{ad}_{c}$. In other words, increasing cell-cell adhesion from low levels 
causes the tumor proliferative capacity to increase. %(representing a mesenchymal phenotype)
We identify an intermediate level of cell-cell adhesion 
where invasiveness and proliferation are maximized. As $f^{ad}$ is 
increased beyond $f^{ad}_{c}$, proliferation capacity is suppressed. 
The non-monotonic dependence of proliferation on $f^{ad}$ qualitatively explains the dual role of 
cell-cell adhesion, as we explain below. By building on the integral feedback mechanism coupling cell dormancy 
and local pressure~\cite{shraiman2005mechanical}, we suggest a physical pressure based formalism for  
the effect of cell-cell adhesion on cell proliferation.  
In particular, we elucidate the role of cell-cell contact, nearest neighbor packing and 
the onset of force dependent cell growth inhibition in influencing 
cell proliferation. 

\section*{Results}
\begin{figure}
 \centering
\includegraphics[width=1\linewidth] {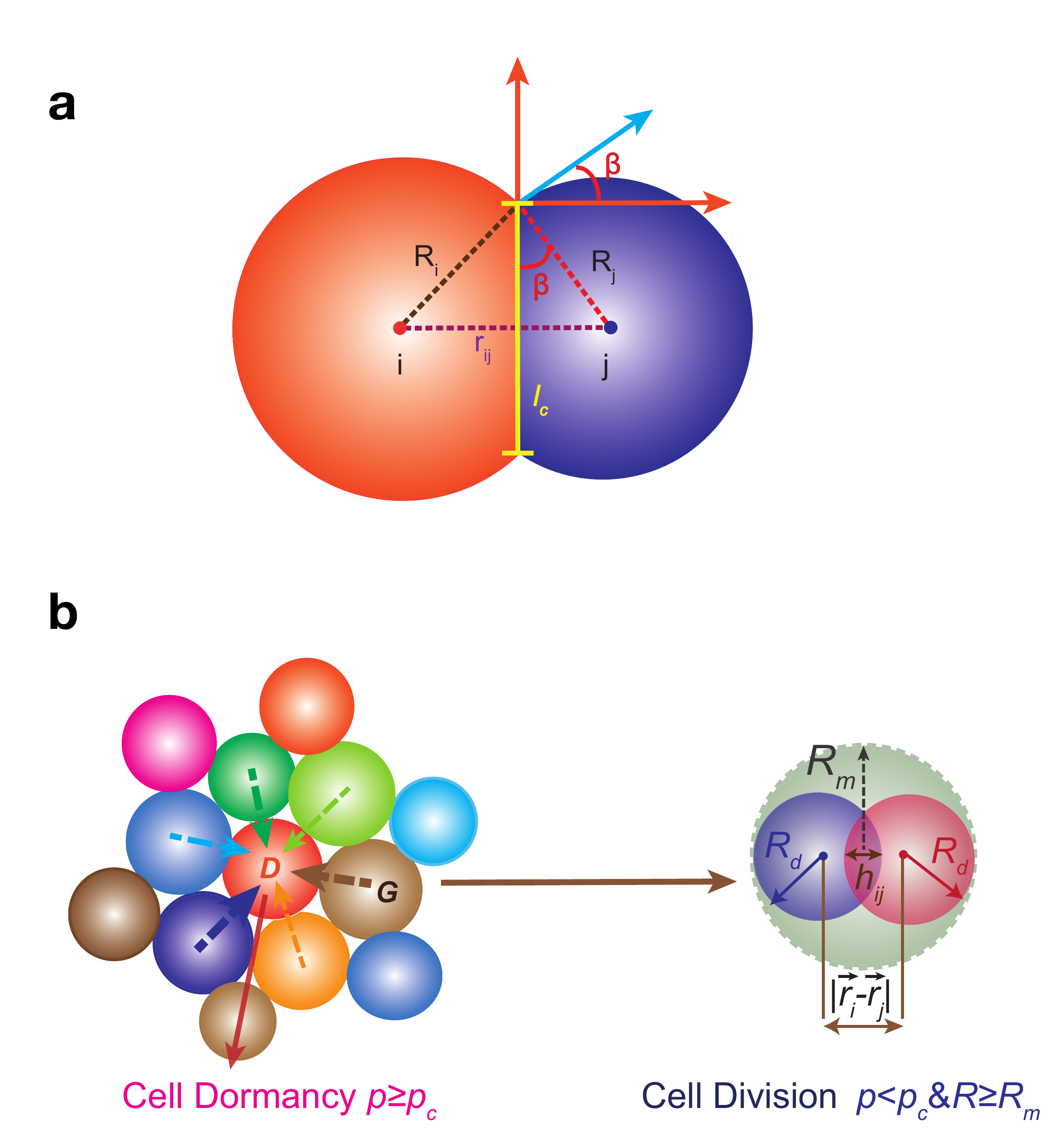}
\caption{{\bf a)} Cell-cell adhesion dictates the angle of contact between cells, $\beta$, and 
the length of contact, ${\it l}_{c}$. 
{\bf b)} Cell dormancy (left) and cell division (right). If the local pressure $p_i$ that the $i^{th}$  cell experiences 
(due to contacts with the neighboring cells) exceeds a specified critical pressure $p_{c}$, it enters 
the dormant state ($D$).  Otherwise, the cells undergo growth (G) until they reach the mitotic radius, $R_{m}$. 
At that stage, the mother cell divides into two identical daughter cells with the same radius $R_{d}$ in such a manner 
that the total volume upon cell division is conserved. A cell that is dormant at a given time can transit from that state at subsequent times. Cell center to center 
distance, $r_{ij}=|\vec{r}_i - \vec{r}_j|$, and cell overlap, $h_{ij}$, are illustrated.}
\label{angle}
\end{figure}

{\bf Cell-Cell Adhesion Strength and Feedback On Cell Proliferation:}
Cell dynamics within proliferating tissues is a complex process,  
where the following contributions at the minimum are coupled:  
(i) cell-cell repulsive and adhesive forces, (ii) cell dynamics 
due to growth, and (iii) cell division and apoptosis. Stochastic cell growth leads to dynamic variations 
in the cell-cell forces while cell division and apoptosis induce temporal rearrangements in the cell positions 
and packing.
%Of particular importance is the role of cell-cell adhesion strength, its effect 
%on force that a cell experiences from its neighbors and how this feeds back into growth and proliferation.  
Cell-cell adhesion strength, dictated by $f^{ad}$ (see Appendix \ref{appforces} Eq.~\ref{ad}), leads to experimentally measurable effects on the 
spatial arrangement of cells (see Fig.~\ref{angle}a), quantified by the angle, $\beta$, and 
the length of contact, ${\it l}_{c}$ between the cells. The angle, $\beta$, should decrease as a function of $f^{ad}$ 
while ${\it l}_{c}$ should increase~\cite{david2014tissue} (see Appendix \ref{appforces} Figs. \ref{forcecomp}a and \ref{forcecomp}b for further details). 

Dynamics of cell-cell contact leads to spatiotemporal fluctuations 
in the local forces experienced by a cell (and vice versa), through which we implement a mechanical 
feedback on growth and proliferation. Depending on the local forces, a cell can either be in the 
dormant ($D$) or in the growth ($G$) phase (see Fig~\ref{angle}b). 
The effect of the local cell microenvironment on proliferation, a collective cell effect, is taken into account through the 
pressure experienced by the cell ($p_i$; see Appendix \ref{appforces}). 
We refer to $p_{i}$ as pressure since it has the same dimensions. 
However, it is not as rigorous a definition of pressure as the thermodynamic conjugate of the volume. 
Essentially, $p_{i}$, models the mechanical sensitivity of cell proliferation to the local environment. 
If $p_{i}$ exceeds $p_c$ (a pre-assigned value of the critical pressure), the cell 
enters dormancy (D), and can no longer grow or divide. However, if $p_{i} < p_{c}$, the cell 
can continue to grow in size until it reaches the mitotic radius $R_m$, the size at which cell division occurs.  

\begin{figure}
 \centering
\includegraphics[width=1\linewidth] {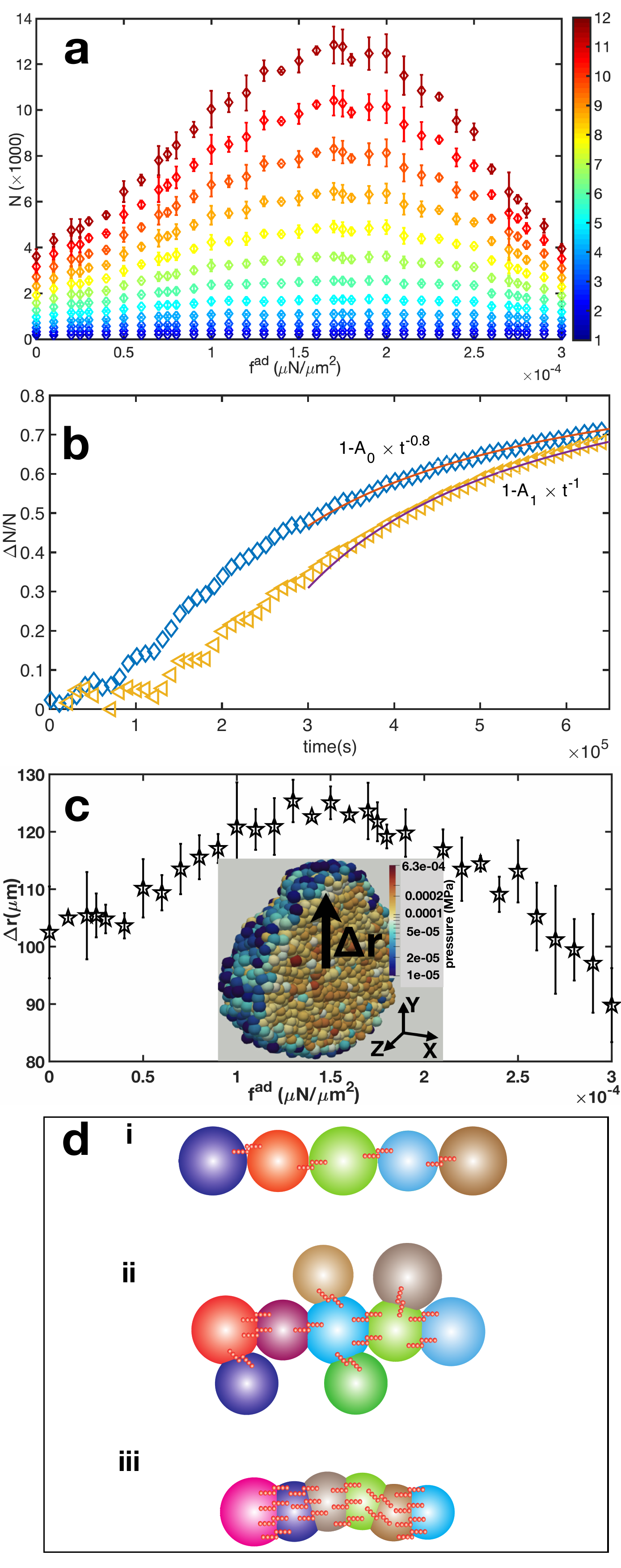}
\caption{Caption next column.}
\label{nvst}
\end{figure}
%\addtocounter{figure}{-1}
%\begin{figure}
%\caption{ }
%\end{figure}
\addtocounter{figure}{-1}
\begin{figure}
\caption{{\bf a)} Proliferation capacity (PrC) measured as 
the total number of cells (N in units of 1000), at $t=\tau_{min}$ to 
$12\tau_{min}$, at intervals of $\tau_{min}(=54,000$ sec), 
as a function of cell-cell adhesion strength ($f^{ad}$) using $p_{c} = 10^{-4}$MPa. 
Error bars here, and henceforth, are calculated from the standard deviation. 
The scale on the right gives $t$ in units of $\tau_{min}$.
{\bf b)} Fractional change in $N$ (defined in the text)
between $f^{ad}=1.75 \times 10^{-4} \mathrm{\mu N/\mu m^2}$ and $f^{ad}=0$ (diamonds), %\mathrm{\mu N/\mu m^2}
between $f^{ad}=1.75 \times 10^{-4}$ and $f^{ad}=3 \times 10^{-4}$ (left triangles),  
as a function of time. Predicted power law behavior of $\Delta N(t)/N$ are plotted as lines. 
{\bf c)} The dependence of the invasion distance ($\Delta r$), at 7.5 
days ($=12\tau_{min}$), on $f^{ad}$. Inset shows a cross section through a tumor spheroid 
and the distance $\Delta r$ from tumor center to periphery at $t\sim 12\tau_{min}$ for $f^{ad}=1.5\times 10^{-4}$. 
Color indicates the pressure experienced by cells. 
{\bf d)} Schematic for the three different regimes of cell-cell adhesion exhibiting differing PrCs. 
E-cadherin molecules are represented as short red bonds. These regimes in the simulations correspond to 
(i) $f^{ad} \le  0.5 \times 10^{-4}$ characterized by low cell-cell adhesion and low PrC. 
(ii) $1\times 10^{-4} \le f^{ad} \le 2 \times 10^{-4}$ characterized by intermediate cell-cell adhesion and high PrC,  
and (iii) $f^{ad} \ge 2.5 \times 10^{-4}$ with high cell-cell adhesion and low PrC.}
\end{figure}

{\bf Proliferation Depends Non-monotonically on {\it f}~$^{ad}$:} 
Our major finding is summarized in Fig.~\ref{nvst}, which shows that tumor proliferation 
and a measure of invasiveness ($\Delta r(t=7.5$~days)), exhibit a non-monotonic dependence on $f^{ad}$. 
Both quantities increase from small values as $f^{ad}$ increases, attain a maximum 
at $f^{ad}=f^{ad}_{c}$, and decrease as $f^{ad}$ exceeds $f^{ad}_{c}$. 
In the simulations, we began with $100$ cells at $t=0$, and let the cells evolve as determined 
by the equations of motion, and the rules governing birth and apoptosis (see Appendix \ref{appforces}). 
We performed simulations at different values of $f^{ad}$ until $\sim 7.5$ days 
($=12\tau_{min}$, where $\tau_{min}$ is the average cell division time), 
sufficient to account for multiple cell division cycles. 
The total number of cells ($N$) as a function of $f^{ad}$ at various times in the range 
of $1\tau_{min}$ to $12\tau_{min}$ (with colors for time $t$ in units of $\tau_{min}$) are shown in Fig.~\ref{nvst}a. 

On increasing $f^{ad}$ from $0$ (no E-cadherin expression) to 
$1.75\times10^{-4}\mathrm{\mu N/\mu m^2}$ (intermediate E-cadherin expression), 
the total number of cells, $N$, at $t=12\tau_{min}$($\sim$7.5 days; dark red in Fig.~\ref{nvst}a) increases substantially.  
When $f^{ad}$ exceeds $f^{ad}_{c}=1.75\times 10^{-4}$, 
the proliferation capacity (PrC) is down-regulated %by $70\%$ 
(Fig.~\ref{nvst}a). While $N=12,000$ cells on day $7.5$ at $f^{ad}=1.75\times10^{-4}\mathrm{\mu N/\mu m^2}$,  
for higher values ($3\times10^{-4}\mathrm{\mu N/\mu m^2}$), 
the tumor consists of only $4,000$ at the same $t$. 
The surprising non-monotonic dependence of cell numbers, $N,$ on $f^{ad}$, 
is qualitatively consistent with some recent experiments~\cite{lewis2010misregulated,baronsky2016reduction,padmanaban2019cadherin}, as we discuss below. The non-monotonic proliferation 
behavior becomes pronounced beginning at $t=5\tau_{min}$ (see Fig.~\ref{nvst}a). The 
fractional change in the number of cells between 
$f^{ad}_{c}$ and another value of $f^{ad}$, $\Delta N(f^{ad},t)/N = [N(f^{ad}_{c}=1.75 \times 10^{-4} \mu N/\mu m^2,t) - N(f^{ad},t)]/N(f^{ad}_{c}=1.75 \times 10^{-4} \mu N/\mu m^2,t)$ 
as a function of time, quantifies the asymmetry in the proliferation capacity due to adhesion strength.  
As shown in Fig.~\ref{nvst}b, the parameter $\Delta N(t)/N$, exhibits non-linear behavior.   
From fits of $N$, as a function of time, the proliferation asymmetry parameter is expected to evolve in time as 
$1 - A_{0} \times t^{-0.8}$ for $f^{ad}=0$ and $1-A_{1}\times t^{-1}$ for $f^{ad}=3 \times 10^{-4} \mu N/\mu m^2$, where 
$A_{0}$ and $A_{1}$ are constants (see Appendix \ref{appforces} Fig. \ref{p1ress}a for the power law fits). 

{\bf Invasion Distance Mirrors Tumor Proliferation Behavior:} The invasion or spreading distance, 
$\Delta r(t)$ (shown in Fig.~\ref{nvst}c), measurable experimentally using imaging methods~\cite{valencia2015collective}, %of the collection of cells 
is the average distance between center of mass of the 
tumor spheroid and the cells at tumor periphery, 
$\Delta r(t) = \frac{1}{N_{b}} \sum_{i}^{N_b} |\vec{r}_{i} - \vec{R}_{CM}|$. 
Here, the summation is over $N_b$, the number of cells at the tumor periphery at 
positions $\vec{r}_{i}$ and $\vec{R}_{CM}$ is the tumor center of mass ($(1/N)\Sigma_{j} \vec{r}_j$). 
In accord with increased proliferation shown in Fig.~\ref{nvst}a,  
$\Delta r(t=12\tau_{min})$ is also enhanced at intermediate 
values of $f^{ad}$ (Fig.~\ref{nvst}c). The uptick in invasiveness 
from low to intermediate values of $f^{ad}$ is fundamentally different 
from what is expected in the canonical picture, where 
increasing cell-cell adhesion suppresses invasiveness and metastatic 
dissemination of cancer cells~\cite{weinberg2013biology,van2008cell}. 
In contrast, tumor invasiveness as a function of 
increasing adhesion or stickiness between cells 
(as tracked by $\Delta r(t=12\tau_{min})$) initially increases and reaches a maximum,  
followed by a crossover to a regime of decreasing invasiveness 
at higher adhesion strengths. 
We note that the decreased invasive behavior at $f^{ad}>f^{ad}_c$ is in 
agreement with the canonical picture, where enhanced E-cadherin expression results in tumor suppression. Schematic 
summary of the results is presented in Fig.~\ref{nvst}d. 
\par
The inset in Fig.~\ref{nvst}c shows the highly heterogenous spatial distribution 
of intercellular pressure (snapshot at $t=12\tau_{min}$), marked by elevated pressure at the core and decreasing 
as one approaches the tissue periphery. As cell rearrangement and birth-death events give rise to local cell density fluctuations, 
the cell pressure is a highly dynamic 
quantity, see videos (Supplementary Movies 1-3; Appendix \ref{appforces} Figs.~\ref{press}a-c) for illustration of pressure dynamics during the growth of the cell collective. 
Spatial distribution of pressure is important to understanding the non-monotonic proliferation behavior, as we discuss in more detail below. 

{\bf Fraction of Dormant Cells Determine Non-monotonic Proliferation:}
To understand the physical factors 
underlying the non-monotonic proliferation behavior shown in Fig.~\ref{nvst}a, we searched for the growth 
control mechanism. %implemented in our study. 
We found that the pressure experienced by a cell as a function 
of cell packing in the 3D spheroid (Appendix \ref{appforces} Fig.~\ref{p1ress}b) plays an essential role in the observed results.  
For $f^{ad}>0$, a minimum in pressure is observed at non-zero cell-cell overlap distance, $h_{ij}$ (Appendix \ref{appforces} Fig.~\ref{p1ress}b). %THIS FIGURE MOVED TO SI see Fig.~\ref{presswi}
For instance, at $f^{ad}=1.5\times 10^{-4} \mathrm{\mu N/\mu m^2 ~ and}~ 3\times 10^{-4} \mathrm{\mu N/\mu m^2}$, 
the pressure ($p_{i}$) experienced by cells is zero at $h_{ij} \sim 0.4\mathrm{\mu m}$ and $1.3\mathrm{\mu m}$ respectively. 
At this minimum pressure, $p_{i} \rightarrow 0$, the proliferation capacity (PrC) of 
the cells is maximized because cells are readily outside the dormant regime, $p_{i}/p_{c} <1$.  
Due to the relationship between cell-cell overlap ($h_{ij}$) and 
center-to-center distance ($|{\mathbf {r}}_{i} -{\mathbf {r}}_{j}|=R_i + R_j -h_{ij}$), 
our conclusions regarding $h_{ij}$ can equivalently be discussed   
in terms of the cell-cell contact length (${\it l}_{c}$), the angle $\beta$ (see Fig.~\ref{angle}) and 
cell-cell internuclear distance, $|r_{i}-r_{j}|$. %Minimum in 

\begin{figure}
 \centering
\includegraphics[width=1\linewidth]{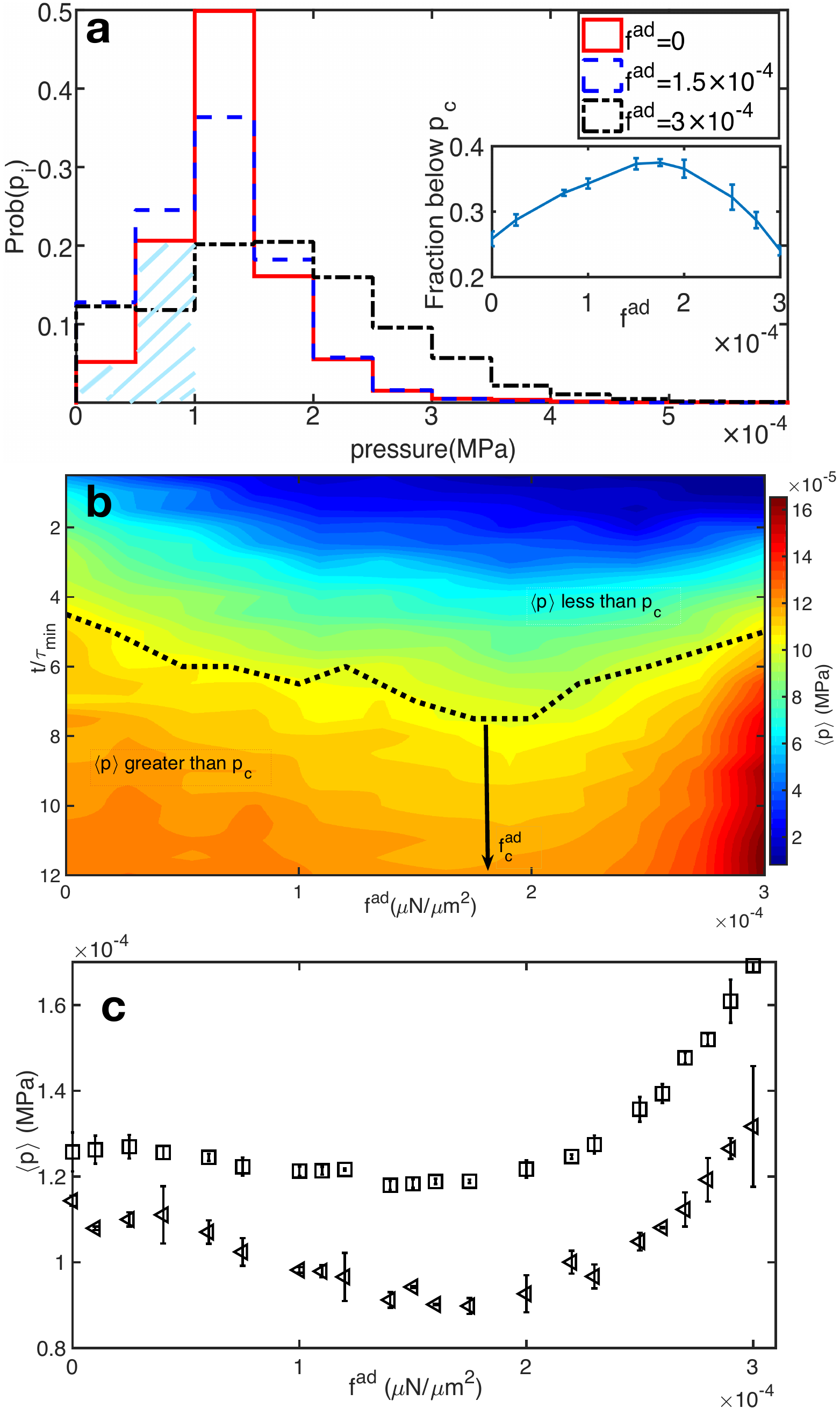} 
\caption{{\bf a)} The probability distribution of pressure experienced 
by cells within a 3D tissue spheroid on day 7.5 for three different 
values of $f^{ad}$. The shaded region represent pressure experienced by cells below 
$p_c=1\times10^{-4}\mathrm{MPa}$, at $f^{ad}=0$. Fraction of cells ($F_{C}$) with 
$p<p_c$ is shown in the Inset. {\bf b)} Phase plot of average pressure experienced by cells as a function of time and $f^{ad}$. The scale for pressure is on the right. 
{\bf c)} The average pressure 
$\langle p \rangle$ experienced by cells at $t=6\tau_{min}$ (left triangles; $\sim$ day 3.75 of growth) and $t=12\tau_{min}$ (squares; $\sim$ day 7.5 of growth) for different values of $f^{ad}$. }
\label{pressdistr}
\end{figure}

Fig.~\ref{pressdistr}a shows the probability distribution of pressure at $t = 7.5~\mathrm{days}$ 
at three representative values of $f^{ad}$, corresponding to low, intermediate and high adhesion strengths. 
The crucial feature that gives rise to the non-monotonic proliferation and invasion (Figs.~\ref{nvst}a -\ref{nvst}c) is the nonlinear 
behavior of the pressure distribution below $p_c=1\times10^{-4}\mathrm{MPa}$ (see shaded portion in 
Fig.~\ref{pressdistr}a, quantified in the Inset) as $f^{ad}$ is changed. 
In the inset of Fig.~\ref{pressdistr}a, the fraction of 
cells in the proliferating regime shows a biphasic behavior with 
a peak at $f^{ad}_{c} \sim 1.75 \times 10^{-4} \mu N/ \mu m^2$. %in agreement with the results in Fig.~\ref{figsum}. 
The fraction of cells in the growth phase ($p_{i} < p_{c}$) peaks at $\approx38\%$, 
between $1\times 10^{-4}~\mathrm{\mu N/\mu m^2} <~f^{ad}~<~2\times 10^{-4}~\mathrm{\mu N/\mu m^2}$. 
For both lower and higher cell-cell adhesion strengths, the fraction of cells in the growth 
phase is at or below $25\%$ on day 7.5 of tissue growth. 
We present the phase diagram of the average pressure ($\langle p \rangle = \frac{\Sigma_i p_i}{N}$, color map) 
on cells as a function of $f^{ad}$ and time (in units of $\tau_{min}$) in Fig.~\ref{pressdistr}b. 
The dotted line marks the boundary $\langle p \rangle=p_{c}$, between the regimes where cells on average 
grow and divide ($ p_{i} < p_{c}$), and the opposite limit where they are dormant. 
At a fixed value of $t$, the regime between 
$1\times 10^{-4} \le f^{ad} \le 2 \times 10^{-4}$ shows a marked dip in the average pressure experienced by cells. 
The low pressure regime, $\langle p \rangle < p_{c}$, is particularly pronounced between 
$2 \tau_{min} \le t \le 7 \tau_{min}$. 
In Fig.~\ref{pressdistr}c, the average pressure 
at $t=6 \tau_{min}$ and $12\tau_{min}$, as a function of $f^{ad}$,  
are shown for illustration purposes. Minimum in $\langle p \rangle$ between 
$f^{ad}\approx1.5-2\times 10^{-4}~\mathrm{\mu N/\mu m^2}$ is seen. 
In our pressure dependent model of contact inhibition, there is a close relationship 
between proliferation capacity (PrC) 
and the local pressure on a cell. As the number of 
cells experiencing pressure below critical pressure increases, 
the PrC of the tissue is enhanced. It is less obvious why the average 
cell pressure acquires a minimum value at $f^{ad}=f^{ad}_{c}$ (Fig.~\ref{pressdistr}b-c). We provide a plausible 
theoretical explanation below. 

{\bf Pressure Gradient Drives Biphasic Proliferation Behavior:}
The finding in Fig.~\ref{nvst}a could be understood using the following physical picture. 
PrC is determined by the number of cells with  
pressures less than $p_{c}$, the critical value. If the pressure on a cell exceeds $p_{c}$, 
it becomes dormant, thus losing the ability to divide and grow. 
The average pressure that a cell experiences depends 
on the magnitude of the net adhesive force. At low values of $f^{ad}$ 
(incrementing from $f^{ad}=0$) the cell pressure  
decreases as the cells overlap with each other because they 
are deformable ($|r_{i} - r_{j}| < R_{i}+R_{j}$). 
This is similar to the pressure in real gases (Van der Waals picture) 
in which the inter-particle attraction leads to a decrease 
in the average pressure. As a result, in a certain range of $f^{ad}$, we expect 
the number of cells capable of dividing should increase, causing enhanced proliferation. At very high values 
of $f^{ad}$, however, the attraction becomes so strong that the number of nearest neighbors of a given cell increases (Appendix \ref{appaverage} Figs.~\ref{figsi2}a-b). 
This leads to an overall increase in pressure (jamming effect), resulting in a decrease in the  
number of cells that can proliferate. From these arguments it follows that $N(t)$ should increase (decrease) at low (high) $f^{ad}$ values 
with a maximum at an intermediate $f^{ad}$. 
The physical picture given above can be used to construct an approximate %mean-field 
theory for the finding that cell proliferation reaches a maximum at $f^{ad} = f^{ad}_{c}$ (Fig.~\ref{nvst}a). 
The total average pressure ($p_{t}$) that a cell experiences is given by 
%\begin{equation}
$p_t \sim \bar{n}_{NN} \langle p_1 \rangle$
%\end{equation} 
where $\bar{n}_{NN}$ is the mean number of nearest neighbors, and 
$\langle p_1 \rangle$ is the average pressure a single neighboring cell exerts on 
a given cell. %(depending on cell-cell overlap $h_{ij}$). 
We appeal to simulations to estimate the dependence of $\bar{n}_{NN}$ %how the average number of nearest neighbors 
on $f^{ad}$ %by numerical fitting %change in $n_{NN}$ to changes in $f^{ad}$ 
(see Appendix \ref{appaverage} Figs.~\ref{figsi2}a-b). %\par
For any cell $i$, the nearest neighbors are defined as those cells with non-zero overlap (i.e. $h_{ij} > 0$). 
To obtain $\langle p_1 \rangle$, we expand around $h_{0}$,
the cell-cell overlap value where both the attractive 
and repulsive interaction terms are equal ($h_{ij}|_{F_{ij}^{el}=F_{ij}^{ad}}$), 
corresponding to the overlap distance at which $p=0$ (Appendix \ref{appforces} Fig.~\ref{p1ress}b). Thus, %Fig.~\ref{presswo
by Taylor expansion to first order, $\langle p_1 \rangle$ can be written as
%\begin{equation}
$\langle p_1 \rangle \approx \frac{\partial p_1}{\partial h} (h-h_{0}) $. 
Here, $h=\bar{h}_{ij}$, is the mean cell-cell overlap, which depends on  
$f^{ad}$ (see Appendix \ref{appaverage} Fig.~\ref{figsi1}). We note that the variation in $h_{0}$ with respect to $R_{i}$ and $R_{j}$,  
as well as other cell-cell interaction parameters is small compared to cell size. 
Estimating the dependence of $h-h_{0}$ 
on adhesion strength (see Appendix \ref{appaverage} Fig.~\ref{figsi3}), an approximate linear trend is observed. 
At higher adhesion strengths, cells find it increasingly difficult to 
rearrange themselves and pack in such a way that intercellular distances are optimal for proliferation.  

In order to calculate the pressure gradient with respect to cell-cell overlap, 
$\frac{\partial p_1}{\partial h}$, we use 
\begin{eqnarray}
\label{pone} 
&&p_1 = \frac{|F|}{A} \simeq \frac{Jh^{\frac{3}{2}} - Bf^{ad}h}{Ch}, \\ %\nonumber 
\label{poneA}
&&\frac{\partial p_1}{\partial h} \simeq \frac{J}{2C h^{\frac{1}{2}}}|_{h=h_{0}} =  \frac{J}{2C h_{0}^{\frac{1}{2}}}, 
\end{eqnarray}
to separate the dependence on cell-cell overlap and adhesion strength. In Eqs.~\ref{pone} and~\ref{poneA} 
$J, B$ and $C$ are independent of both $h$ and $f^{ad}$ and can be obtained from  
Appendix \ref{appforces} Eqs.~\ref{rep} and ~\ref{ad}, and the definition of $A_{ij}$. 
The resulting expressions are: $J=1/[{\frac{3}{4}(\frac{1-\nu_{i}^2}{E_i} + \frac{1-\nu_{j}^2}{E_j})\sqrt{\frac{1}{R_{i}(t)}+ \frac{1}{R_{j}(t)}}}], 
B=[\pi R_{i}R_{j}/(R_{i}+R_{j})]\times \frac{1}{2}(c_{i}^{rec}c_{j}^{lig} + c_{j}^{rec}c_{i}^{lig}), ~\mathrm{and} ~C=\pi R_{i}R_{j}/(R_{i}+R_{j})$. 
On equating the repulsive and attractive interaction terms, we 
obtain $h_{0} \approx K(f^{ad})^2 $, implying 
$\frac{\partial p_1}{\partial h} = \frac{J}{2C} \frac{1}{\sqrt{K} f^{ad}}$, where, 
$K=(B/J)^{2}$. 
Thus, the total pressure ($p_t$) experienced by a cell is $p_t \sim \bar{n}_{NN} \langle p_1 \rangle = \bar{n}_{NN} (h-h_{0}) \frac{\partial p_1}{\partial h}$, 
%\begin{equation}
%\begin{eqnarray}
%p_t &&\sim n_{NN} \langle p_1 \rangle, \\
%      &&= n_{NN} (h-h_{o}) \frac{\partial p_1}{\partial h}, \\
%      &&=(G + \beta f^{ad})(E + \alpha f^{ad})(\frac{J}{2C}\frac{1}{\sqrt{K} f^{ad}}). \label{ptheory}
%\end{eqnarray}
\begin{equation}
p_t =(G + \beta f^{ad})(E + \alpha f^{ad})(\frac{J}{2C}\frac{1}{\sqrt{K} f^{ad}}). \label{ptheory}
\end{equation}

The mean number of near-neighbors $\bar{n}_{NN}$ increases with $f^{ad}$ and to a 
first approximation can be written as $G + \beta f^{ad}$ (see Appendix \ref{appaverage} Fig.~\ref{figsi2}a-b; $G,\beta$ are constants obtained from fitting simulation data). Similarly, 
the deviation of the cell-cell overlap from $h_{0}$ is approximately $E + \alpha f^{ad}$ (see Appendix \ref{appaverage} Fig.~\ref{figsi3}; $E,\alpha$ are constants). 
%\end{equation}
Notice that Eq.~\ref{ptheory} can be written as, %cast as, 
%\begin{equation}
$\Omega p_{t} = (G\alpha + \beta E) + \beta \alpha f^{ad} + \frac{GE}{f^{ad}}$,
%\end{equation}
where $\Omega=(\frac{J}{2C\sqrt{K}})^{-1}$. In this form, the second term depends linearly on $f^{ad}$ and 
the third is inversely proportional to $f^{ad}$. As described in the physical arguments, enhancement in proliferation is maximized 
if $p_{t}$ is as small as possible. 
The minimum in the total pressure experienced is given by the solution to 
$\frac{\partial p_t}{\partial f^{ad}}=0$. 
Therefore, the predicted optimal cell-cell adhesion strength is  
$f^{ad}_{opt}=(\frac{GE}{\alpha \beta})^\frac{1}{2}= 1.77 \times 10^{-4} \mathrm{\mu N/\mu m^2}$. 
This is in excellent agreement with the simulation results ($f^{ad}_{c} \approx 1.75 \times 10^{-4} \mu N/ \mu m^2$) for the peak in the proliferation 
behavior $(N(t=7.5~\mathrm{days})$ in Fig.~\ref{nvst}a). More importantly, the arguments leading to Eq.~\ref{ptheory} 
show that the variations in the average pressure as a function of $f^{ad}$ drives proliferation. 
%\bigskip

\begin{figure}
 \centering
\includegraphics[width=1\linewidth]{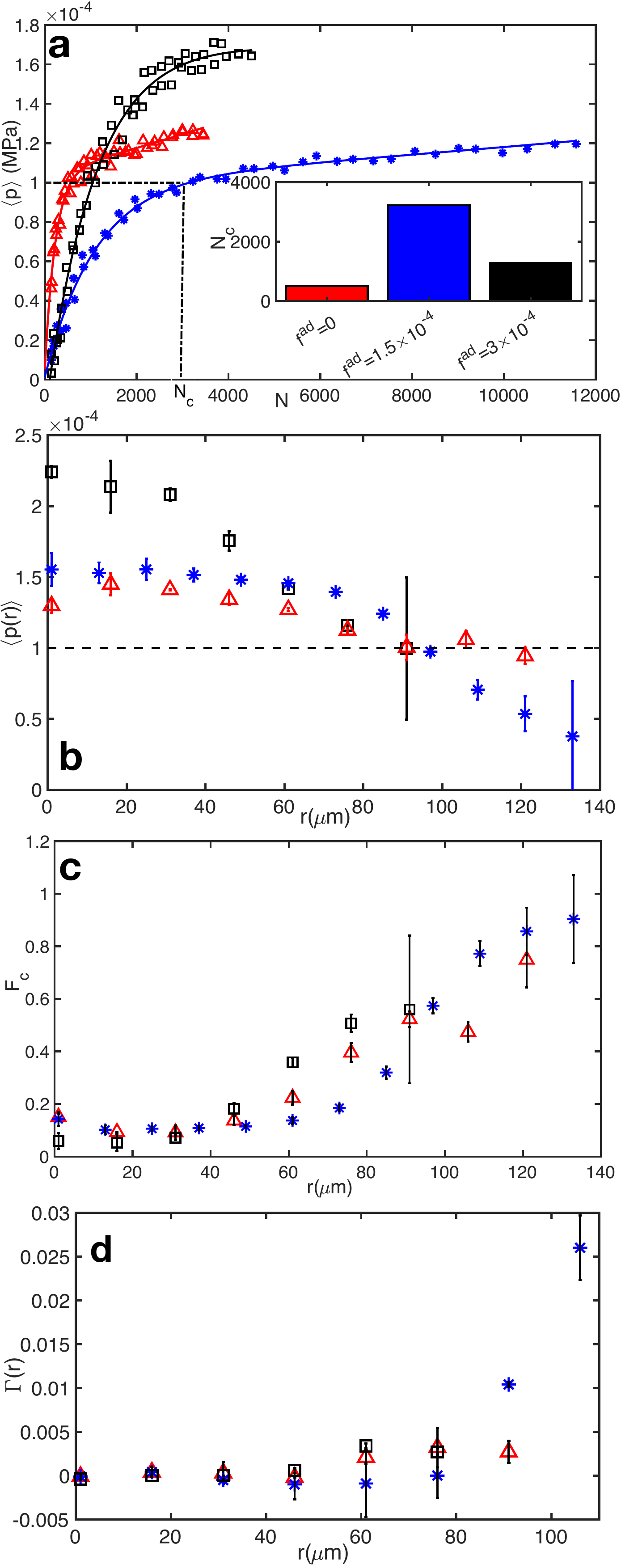}
\caption{(Caption next column.)}
\label{eos}
\end{figure}

{\bf Cell Packing and Spatial Proliferation Patterns Are Dictated by Cell-Cell Adhesion:} 
%A thermodynamic equation of state for active matter, such as the relation between %change3
%pressure and density of an ideal gas, is as of yet an unresolved question~\cite{cugliandolo2011effective,ginot2015nonequilibrium,bartolo2017viewpoint}. 
Mechanosensitivity i.e. specialized response to mechanical stimulation is common to cells in many organisms. 
Exposure to stresses in living tissues can module physiological processes from molecular, to cellular and systemic levels~\cite{kamkin2005mechanosensitivity}. 
Here, we quantify the spatio-temporal behavior of the intercellular pressure, the parameter 
that encodes the mechanical sensitivity of cell proliferation to the local environment, and 
delineate its emergent properties in a growing cell collective. 
The average pressure, $\langle p \rangle$, experienced by cells as a function of the 
total number of cells at varying $f^{ad}$ is shown in Fig~\ref{eos}a. 
$N_{c}$, defined as the number of cells at which $\langle p \rangle = p_{c}$, exhibits 
a biphasic behavior, supporting the maximum number of cells at an intermediate $f^{ad}$. This provides further evidence that 
in a growing collection of cells, at intermediate $f^{ad}$,  
cells rearrange and pack effectively in such a manner that the average pressure is minimized. 
For all three adhesion strengths considered, 
an initial regime where pressure rises rapidly is followed by a more 
gradual increase in pressure, coinciding with the exponential to power law crossover in the growth in the 
number of cells (see Fig.~\ref{p1ress}a in Appendix \ref{appforces}). 
$\langle p \rangle$ as a function of $N$ are well fit 
by double exponential functions. 
We propose that the intercellular pressure in 3D cell collectives 
should exhibit a double exponential dependence on $N$. However, the precise nature of the intercellular pressure 
depends on the details of the interaction between cells. 
\par
%\section{Comparison to Experiments}
With recent advances in experimental techniques~\cite{dolega2017cell,mongera2018fluid}, it is now possible to map spatial variations in 
intercellular forces within 3D tissues. 
Hence, we study how the spatial distribution of pressure and proliferation is influenced by cell-cell adhesion strength. 
We find that the cells at the tumor center experience higher pressures, above the 
critical pressure $p_{c}$ (see Fig.~\ref{eos}b), independent of $f^{ad}$. 
%At larger values of $f^{ad}$ we find that higher average pressure
%is observed for cells at the smaller distances of $r$ from the tumor center. 
In contrast, the average pressure experienced by  
cells close to the tumor periphery is below the critical value $p_{c}$. The pressure 
decreases as a function of distance $r$ from the tumor center, 
with the lowest average pressure observed at the intermediate value of 
$f^{ad}= 1.5\times 10^{-4}$ as one approaches the tumor periphery. 
We calculate the average pressure dependence on $r$ using,  
\begin{equation}
\langle p(r) \rangle = \frac{\Sigma_{i} p_{i}\delta[r-(|\vec{R}_{CM} -\vec{r}_{i})]}{\Sigma_{i}\delta[r-(|\vec{R}_{CM} -\vec{r}_{i})]}.
\end{equation}
Due to the high pressure experienced by 
cells near the tumor center, a low fraction ($F_{c}<0.2$) of cells are in growth 
phase at small $r < 50\mu m$ while the majority of cells can grow at large $r$ 
(see Fig.~\ref{eos}c).  
A rapid increase in $F_{c}$ is observed approaching 
the tumor periphery for the intermediate value of $f^{ad}= 1.5\times 10^{-4}$ 
(see the blue asterisks in Fig.~\ref{eos}c). To understand the rapid tumor invasion at 
intermediate value of $f^{ad}$, we calculated the average cell proliferation rate, $\Gamma(r)$, %(see SI Section VII for more details)). 
at distance $r$ from the tumor center,
$\Gamma(r) = \frac{N(r,t)-N(r,t-\delta t)}{\delta t}.$ 
Here, $N(r,t)$ is the number of cells at time $t=650,000s$ and $\delta t=5000s$ is the time interval. 
The average is over polar and azimuthal angles, for all cells between $r$ to $r+\delta r$. 
%$\Gamma(r)$ is calculated at $t=650,000s$ and $\delta t=5000s$.  
Closer to the tumor center, at low $r$, $\Gamma (r)\sim 0$ indicating no proliferative activity. 
However, for larger values of $r$, proliferation rate rapidly increases approaching the periphery. 
%The proliferation rate is highest for $f^{ad}=1.5\times 10^{-4}$, approaching the periphery.
We found that the cell proliferation rate is similar for different 
$f^{ad}$ at small $r$, while a much higher proliferation rate is observed 
for the intermediate value of $f^{ad}$ at larger $r$ (see Fig.~\ref{eos}d).
This spatial proliferation profile is in agreement with experimental results, where increased 
mechanical stress is correlated with lack of proliferation within the spheroid core, albeit in the context 
of externally applied stress~\cite{dolega2017cell}. We show, however, that even in the absence 
of an external applied stress, intercellular interactions give rise heterogeneity in the spatial 
distribution of mechanical stresses.

\section*{Discussion}

{\bf Internal Pressure Provides Feedback In E-cadherin's Role In Tumor Dynamics:} 
We have shown that the growth and invasiveness of the tumor spheroid changes non-monotonically with 
$f^{ad}$, exhibiting a maximum at an intermediate value of the inter-cell adhesion strength. The mechanism 
for this unexpected finding is related to a collective effect that alters the pressure on a cell 
due to its neighbors. %, which maybe thought of as a jamming effect. 
Thus, internal pressure may be viewed as 
providing a feedback mechanism in tumor growth and inhibition. 
The optimal value of $f^{ad}_{c}$ 
at which cell proliferation is a maximum at $t~ >> \tau_{min}$, is due to  
$p_{c}$, the critical pressure above which a cell enters dormancy. Taken together these results 
show that the observed non-monotonic behavior is due to an interplay of $f^{ad}$ and pressure, 
which serves as a feedback in enhancing or suppressing tumor growth. 
We note that the main conclusion of our model, on the non-monotonic proliferation behavior with a maximum at $f^{ad} \approx f_c^{ad}$, 
is independent of the exact value of $p_c$ (see SI Fig. S1), 
alternative definitions of pressure experienced by the cells as well as the cell-cell interaction (see SI Figs. S2-S3; see SI Section I and II for more details).

\addtocounter{figure}{-1}
\begin{figure}
\caption{{\bf a)} (Figure in previous page) Average pressure experienced by cells, $\langle p \rangle$, as a function of 
the total number, $N$, of cells; $\langle p \rangle$ versus $N$ at $f^{ad}=0$ (red;triangle), $f^{ad}=1.5 \times 10^{-4}$ (blue;asterisk) and 
$f^{ad}=3 \times 10^{-4}$ (black;squares) are shown. The corresponding double exponential fits are, 
[$1.1\times 10^{-4} e^{5.8\times10^{-5}\times N}-1\times 10^{-4} e^{-4.4\times 10^{-3}\times N}$] (red line, $f^{ad}=0$), 
[$1.0\times 10^{-4} e^{1.7\times10^{-5}\times N}-9.9\times 10^{-5} e^{-9.2\times 10^{-4}\times N}$] (blue line, $f^{ad}=1.5 \times 10^{-4}$) 
and [$1.8\times 10^{-4} e^{7.6\times10^{-6}\times N}-1.9\times 10^{-4} e^{-9.1\times 10^{-4}\times N}$] (black line, $f^{ad}=3 \times 10^{-4}$). Inset shows $N_{c}$, the number of cells at which $\langle p \rangle = p_{c}$.
{\bf b)} The average pressure experienced by cells at a distance $r$ ($\mathrm{\mu m}$) from the spheroid center. %for $f^{ad}=0$ (red), 
%$f^{ad}=1.5\times 10^{-4}$ (blue) and $f^{ad}=3\times 10^{-4}$ (black). 
The dashed line shows the critical pressure $p_{c} = 10^{-4}\mathrm{MPa}$.  
{\bf c)} The fraction of cells with $p < p_{c}$ at a distance $r$. %for $f^{ad}=0$ (blue), 
%$f^{ad}=1.5\times 10^{-4}$ (red) and $f^{ad}=3\times 10^{-4}$ (yellow). 
{\bf d)} The average cell proliferation rate at distance $r$ from the center of the spheroid for $f^{ad}=0$, 
$f^{ad}=1.5\times 10^{-4}$ and $f^{ad}=3\times 10^{-4}$. Colors and symbols corresponding to $f^{ad}$ 
are the same for {\bf a}-{\bf d}. $t$ is fixed at $650,000sec (\sim 12\tau_{min})$ for {\bf b}-{\bf d}.}
\end{figure}

The growth mechanism leading to non-monotonic proliferation at times exceeding a 
few cell division cycles is determined by the fraction of cells, $F_C$, with pressure 
less than $p_c$. The growth rate, and hence $F_C$, depends on both $p_c$ as well as $f^{ad}$. 
This picture, arising in our simulations, is very similar to the mechanical feedback as a control mechanism for tissue 
growth proposed by Shraiman~\cite{shraiman2005mechanical,irvine2017mechanical}. In his formulation, 
the tissue is densely packed (perhaps confluent) so that cellular rearrangements 
does not occur readily, and the tissue could be treated as an elastic sheet that 
resists shear. For this case, Shraiman produced a theory for uniform tissue growth 
by proposing that mechanical stresses serves as a feedback mechanism. In 
our case, large scale cell rearrangements are possible as a cell or group of cells could
go in and out of dormancy, determined by the $p_{i}(t)/p_{c}$. %where $p_i$ is the pressure on the $i^{th}$ cell at time $t$. 
Despite the differences between the 
two studies, the idea that pressure could serve as a regulatory mechanism of 
growth, which in our case leads to non-monotonic dependence of proliferation 
on cell-cell adhesive interaction strength, could be a general feature of mechanical feedback\cite{shraiman2005mechanical}. 

\begin{figure}
 \centering
\includegraphics[width=0.95\linewidth]{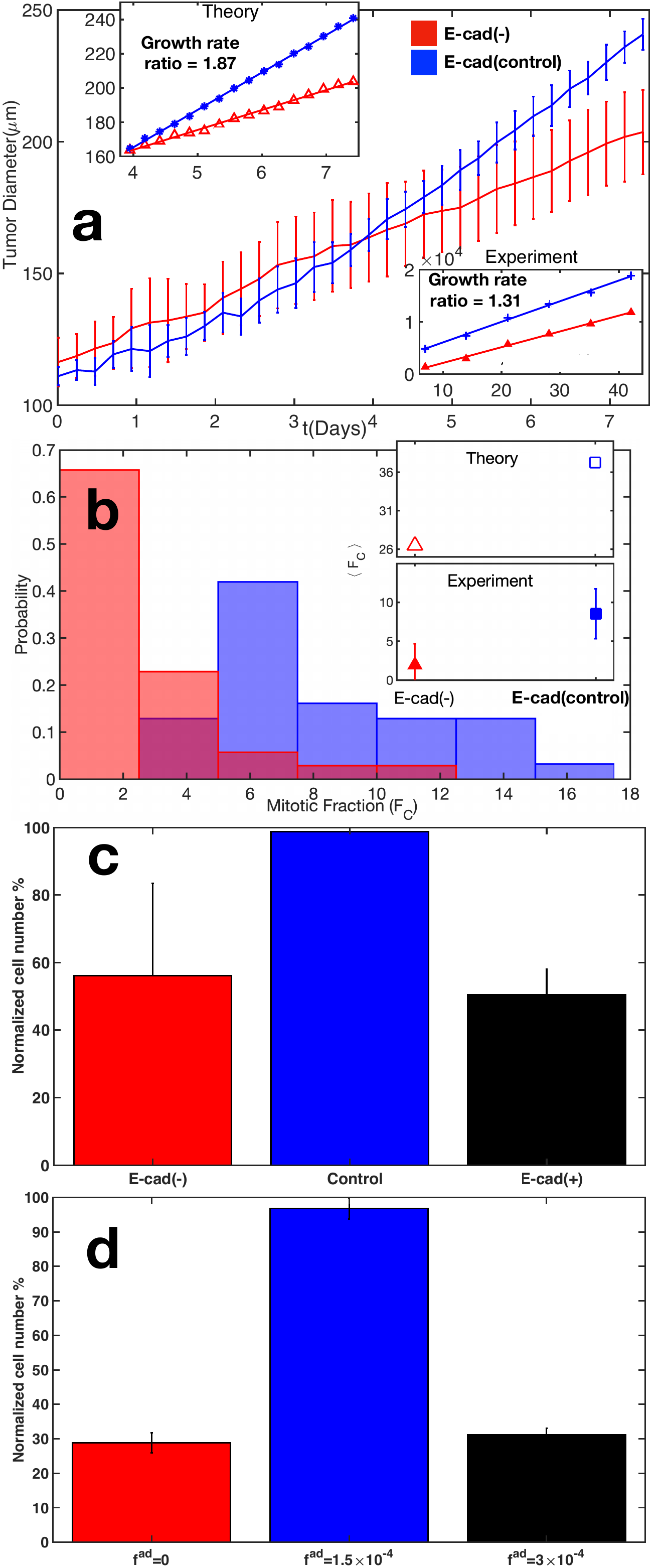}
%\sidesubfloat[]{\includegraphics[width=0.53\linewidth] {cell_num_Ecad_latest2.eps} \label{expt}}
\caption{(Caption next column.)}
\label{fig5}
\end{figure}
\addtocounter{figure}{-1}
\begin{figure}
\caption{{\bf a)} Main Panel) Kinetics of growth in the diameter of tumor spheroids 
composed of cells with low E-cadherin expression (red; $f^{ad}=0$; E-cad(-)) 
and intermediate E-cadherin expression (blue; $f^{ad}=1.75 \times 10^{-4}$; E-cad(control)) from simulations show 
enhanced tumor growth rate due to higher E-cadherin expression. The tumor diameter growth is linear 
at $t>4$ days, independent of the E-cadherin expression level. However, the growth rate of the tumor colony with 
intermediate E-cadherin expression is larger (Top Inset; Theory) in agreement with experimental results. 
Growth rates of the longest tumor dimension 
in low E-cadherin expressing organoids compared to control organoids with normal E-cadherin expression 
is shown in the Bottom Inset (Data was extracted from Ref.~\cite{padmanaban2019cadherin}). 
{\bf b)} Probability distribution of the mitotic fraction, $F_C$, in cell colonies expressed as $\%$ 
for low E-cadherin tumor organoids (red) and normal E-cadherin expressing tumor organoids (blue) 
shows enhanced proliferation capacity when E-cadherin expression is increased. Data was 
extracted from Ref.~\cite{padmanaban2019cadherin}. 
(Top Inset; Theory) Mean mitotic fraction, $\langle F_C \rangle$, for low E-cadherin tumor cells (E-cad(-)) 
and intermediate E-cadherin expressing tumor cells (E-cad (control)). 
(Bottom Inset; Experiment) Mean mitotic fraction in low E-cadherin expressing tumor organoids is lower 
than control organoids with normal E-cadherin expression~\cite{padmanaban2019cadherin}. 
Although there is paucity of experimental data it is encouraging that the currently available measurements are in line with 
our predictions.
{\bf c)} Experimental data for the number of 
cells normalized by maximum number (as $\%$) at three levels of E-cadherin expression 
for primordial germ cells (PGC) from 
{\it Xenopus laevis} $\approx$1.5 days after ferilization~\cite{baronsky2016reduction}. 
Both overexpression and knockdown of 
E-cadherin leads to a decrease in PGC cell numbers. %, in fair agreement with 
%the simulation results. %Data from Western Blot analysis in Ref.~\cite{baronsky2016reduction}
%is presented in the inset. %Embryos were injected with 
Control morpholino oligonucleotides (Contr MO) and  
%E-cadherin morpholino oligonucleotides (E-cad MO) or
%GFP E-cadherin DELE mRNA (E-cad GFP). 
uninjected show native E-cadherin expression while 
E-cad MO (E-cad(-)) and E-cad GFP (E-cad(+)) induces knockdown and 
overexpression of E-cadherin respectively. 
{\bf d)} Simulation data for the number of 
cells normalized by maximum number (as $\%$) at three levels of cell-cell adhesion strength ($t=650,000sec (\sim 12\tau_{min})$). 
Values of $f^{ad}$ (in units of $\mu N/\mu m^{2}$) are shown in parenthesis, 
as a proxy for experimental E-cadherin expression levels.}
\end{figure}

{\bf Cell-matrix Interactions:} Biphasic cell migration controlled by adhesion to the extracellular 
matrix has been previously established. In pioneering studies, Lauffenburger and 
coworkers~\cite{dimilla1991mathematical,dimilla1993maximal} showed using theory and 
experiments that speed of single cell migration is biphasic, achieving an optimal value at an intermediate 
strength of cell-substratum interactions. Similarly, invasion of melanoma cells into a matrix was also 
found to be biphasic \cite{ahmadzadeh2017modeling}, increasing at small collagen concentrations and reaching a peak at an intermediate value. 
At much higher values of the collagen concentration, the invasion into the matrix decreases giving rise to the biphasic dependence. A direct link 
between the survival of genetically induced glioma-bearing mice and the expression level of CD44, 
which is a cell surface marker, has recently been reported \cite{klank2017biphasic}. The authors 
showed that the survival depends in a biphasic manner with increasing CD44 expression. Simulations using the motor clutch 
model established that the results could be explained in terms of the strength of cell-substrate interactions. 
In contrast to these studies, our results show that adhesion strength between cells, mediated by E-cadherin expression, 
gives rise to the observed non-monotonic behavior in the tumor proliferation (Fig.~\ref{nvst}a). 
The mechanism, identified here, is related to the pressure dependent feedback on growth~\cite{shraiman2005mechanical}  
whose effectiveness is controlled by cell-cell adhesion strength, $f^{ad}$.  

{\bf Comparison With Experiments:} We consider three experiments, which provide 
support to our conclusions: (i) Based on the observation of enhanced cell migration due to 
lower E-cadherin levels, it has been proposed that tumor invasion and metastasis follows 
the loss of E-cadherin expression~\cite{frixen1991cadherin}. However, most breast cancer 
primary and metastatic tumors express E-cadherin~\cite{li2003trends}. To resolve this discrepancy, 
Padmanaban et. al~\cite{padmanaban2019cadherin} compared the tumor growth behavior 
between E-cadherin negative cells (characterized by reduced E-cadherin expression compared to control) and control E-cadherin expressing cells using three-dimensional (3D) tumor organoids. 
They report that tumors arising from low E-cadherin expressing cells (E-cad(-)) were smaller than 
tumors from control E-cadherin (E-cad(control)) expressing cells at corresponding time points over multiple weeks of 
tumor growth (see Fig.~\ref{fig5}a Lower Inset). 
Using the experimental data from Ref.~\cite{padmanaban2019cadherin}, we extracted the tumor growth rate for    
E-cad(-) and E-cad(control) organoids, and observe enhanced growth rate for tumors made up of E-cad(control) cells. 
%The longest tumor dimension expanded at a rate of $395~\mathrm{\mu m/day}$ for E-cadh(control) tumor cells while 
%E-cad(-) tumor organoid expanded at a rate of $302~\mathrm{\mu m/day}$ (see Bottom Inset, Fig.~\ref{fig5}a). 
The longest tumor dimension for E-cadh(control) tumor organoids expanded $1.3$ times faster compared to  
E-cad(-) tumor organoids (see Bottom Inset, Fig.~\ref{fig5}a). 
From our tumor simulations, we predicted the E-cadh(control) tumor spheroids expand $1.8$ times faster compared to 
E-cad(-) tumor spheroids (see Top Inset, Fig.~\ref{fig5}a). 
This shows good agreement between theory and experimental results. 
It is worth emphasizing that the simulation results were obtained without any fits to the experiments~\cite{padmanaban2019cadherin}, 
which appeared while this article was already submitted. 
Comparison of the growth in tumor diameter between E-cad(-) cells 
and E-cad(control) (red and blue lines respectively, Fig.~\ref{fig5}a Main Panel) over $7.5$ days of simulated tumor growth shows similar growth rates 
until $t\sim 4$ days followed by enhanced growth rates for E-cad(control) cells at $t> 4$ days. 
The fit for the simulated tumor diameter growth rate is obtained by analyzing data at $t>4$ days (see Top Inset, Fig.~\ref{fig5}a). 
The tumor size growth rate over multiple days is best fit by a linear function, in agreement between simulation and experiment. 
%Our model prediction for the enhancement in tumor growth rate in E-cad(control) cells as compared to E-cad(-) cells is observed in experiments. 
The overall magnitude of the tumor growth rate is higher in the experiments as compared to the simulations.  
The difference in the magnitude of tumor growth rate could be due to the variation in the sizes of cells. In simulations, the maximum cell diameter is $10~\mu m$ while cells can be larger 
than $20~\mu m$ in the experimental tumor spheroids. 
Other factors such as the difference in cell cycle times, critical pressure which limits growth and spatially asymmetric growth of the tumor could 
also lead to differences in the overall magnitude of the tumor growth rate. 
However, both the linear functional form of the tumor growth rates and the higher growth rates in E-cad(control) tumor spheroids 
as compared to E-cad(-) tumor spheroids are in agreement with simulation predictions.

To conclusively show that the fraction of proliferating cells determine the non-monotonic tumor 
growth behavior between low E-cadherin expressing tumors and cells expressing control levels of E-cadherin, we turn to further analysis of  
experimental data from Ref.~\cite{padmanaban2019cadherin}. 
By staining tumor cell colonies for PH3 (phospho-histone 3; a mitotic marker), Ref.~\cite{padmanaban2019cadherin} 
quantifies the mitotic fraction i.e. the ratio of the number of actively dividing cells to the total number of cells. 
In agreement with predictions from our simulations, Padmanaban et. al~\cite{padmanaban2019cadherin} report that  
E-cadherin expressing tumor cell colonies are indeed characterized by a larger mitotic fraction (blue histogram, Fig.~\ref{fig5}b) 
as opposed to low E-cadherin expressing tumor cells (red histogram, Fig.~\ref{fig5}b). We compare the mean 
mitotic fraction, $\langle F_{C} \rangle$, in E-cad(-) and E-cad(control) in the insets of Fig.~\ref{fig5}b between theory and experiment. 
In agreement with our simulation predictions, enhanced mitotic fraction is seen in 
tumor spheroids with normal E-cad expression (see Insets of Fig.~\ref{fig5}b).

(ii) Maintenance of appropriate E-cadherin expression is required for normal 
cell development in several species such as drosophila, zebrafish and mouse~\cite{baronsky2016reduction,richardson2010mechanisms}. 
%In order to make quantitative connections between our findings and experiments, 
%examining the consequence of interfering with expression levels of E-cadherin, %in the context of embryogenesis, 
We consider a study on how the proliferation capacity of primordial germ cells (PGCs) 
depend on cell-cell adhesion. Even though the total 
number of PGCs in an embryo is only around 10-30 cells, while there are thousands of cells in the simulations, 
we compare the percentage change in cell 
numbers that result from modulating E-cadherin expression levels. We make this comparison 
to propose that perhaps E-cadherin does modulate proliferation through a pressure dependent 
mechanistic pathway in tissues, as discussed above. 
The proliferation behavior of PGCs in 
{\it Xenopus laevis} %(at stages 28-30) 
with changing E-cadherin expression levels is shown in Fig.~\ref{fig5}c. In this experiment, 
E-cadherin overexpression, achieved by mRNA injection of GFP E-cadherin DELE mRNA (E-cad GFP), 
leads to $\sim 50\%$ reduction (compared to control) in the number of cells 
after $1.5$ days of post fertilization embryo growth~\cite{baronsky2016reduction}. 
Similar reduction in the number of cells is observed with E-cadherin 
knockdown using specific morpholino oligonucleotides (E-cad MO)~\cite{baronsky2016reduction}. 
Therefore, an optimal level of cell-cell adhesion exists where proliferation is maximized, in agreement 
with simulation predictions (Fig.~\ref{fig5}d). The section III in SI provides 
other examples for the role of E-cadherin in both tumor suppression and proliferation. 
(iii) The loss of E-cadherin is considered 
to be a key characteristic in epithelial to mesenchymal transitions, 
priming the cells to dissociate from the primary tumor and invade surrounding tissues~\cite{mendonsa2018ecadherin}. 
%However, the effect of E-cadherin as a tumor promoter is not clearly understood. 
However, in a subset of high grade glioblastoma, patients with tumor cells expressing E-cadherin correlated 
with worse prognosis compared to patients whose tumor cells that did not express E-cadherin~\cite{lewis2010misregulated}. 
In this tumor type, heightened expression of E-cadherin correlated with increased invasiveness 
in xenograft models~\cite{lewis2010misregulated}. These experimental results, which are consistent with our simulations in 
promoting proliferation as $f^{ad}$ is changed from $=0~\mathrm{\mu N/\mu m^2}$ to $f^{ad}=1.75 ~\mathrm{\mu N/\mu m^2}=f^{ad}_{c}$ (Fig.~\ref{nvst}a), 
suggest an unexpected role of E-cadherin in promoting tumor growth and invasion. 

{\bf Cautionary Remarks:} As detailed in the SI section III, it is difficult to make precise comparisons between 
the simulations and experiments because the growth of tumors is extremely complicated.  
For example, other cytoplasmic and nuclear signaling factors 
may be important in understanding the role of cell-cell adhesion in modulating 
the proliferative capacity of cells,~\cite{rodriguez2012cadherin} as multiple signaling pathways are located in direct 
proximity to the adherens junction complexes~\cite{kourtidis2017central}. 
Nevertheless, our results suggest that the mechanism of contact 
inhibition of proliferation, based on critical cellular pressure,   
could serve as a unifying mechanism in understanding how cell-cell adhesion 
influences proliferation. 
The relation between proliferation and cell-cell adhesion has important 
clinical applications, such as in the development of innovative therapeutic approaches to 
cancer~\cite{brouxhon2013monoclonal,carneiro2013therapeutic} by targeting E-cadherin expression. 
%Given the interest in E-cadherin modulation as a therapeutic strategy for cancer, it is important to consider 
%the biphasic cell proliferation behavior we have identified. 
Under certain circumstances, inhibiting E-cadherin expression could lead to tumor progression. 
On the other hand, in cancer cells nominally associated with low or negligible E-Cadherin levels, 
tumor progression and worsening prognosis could result upon increasing E-Cadherin levels (see SI, Section III for further discussion).
%Therefore, 
%it is likely that tumor stratification based on high, intermediate and low E-cadherin expression 
%maybe useful as a treatment strategy (see SI, Section III for further discussion). 
%A similar therapeutic strategy was recommended based on $in~vivo$ experiments validating the biphasic dependence 
%of cell migration on CD44 levels~\cite{klank2017biphasic}.   Although a direct connection to specific cancers using our results 
%cannot be made, we provide partial evidence for the counterintuitive experimental finding that enhanced 
%adhesion between cells could promote cell proliferation. 

\section*{Conclusions} 
In this study, we have established that %find that a simple monotonic description of 
the modulation of cell-cell adhesion strength %mediated by E-cadherin expression, %on cell proliferation confuses critical aspects of how E-cadherin 
contributes to contact inhibition of cell growth and proliferation. Surprisingly, %a non-monotonic 
cell proliferation exhibits a non-monotonic behavior as a function of cell adhesion strength, increasing till a critical value, 
followed by proliferation suppression at higher values of $f^{ad}$. We have shown that E-cadherin expression  
and critical pressure based contact inhibition 
are sufficient to explain the role of cell-cell adhesion on cell proliferation in the context of both 
morphogenesis and cancer progression. The observed dual role that E-cadherin plays in tumor growth  
is related to a feedback mechanism due to changes in pressure as the cell-cell interaction strength is varied, established here 
on the basis of simulations and a mean field theory. 

The pressure feedback on the growth of cells is sufficient to account for cell proliferation in the simulations. 
For cells, however, it may well be that mechanical forces do not directly translate into proliferative effects. Rather, 
cell-cell contact (experimentally measurable through the contact length $l_{c}$, for example) could biochemically regulate Rac1/RhoA signaling, which 
in turn controls proliferation, as observed in biphasic proliferation of cell collectives in both two and three dimensions~\cite{liu2006cadherin,gray2008engineering}. 

One implication of our finding is that the mechanical pressure dependent feedback may also play a role in organ size control.   
As tissue size regulation requires fine tuning of proliferation rate, cell volume, and cell death at the single cell level~\cite{levayer2016tissue}, 
pressure dependent feedback mediated by cell-cell adhesion could function as an efficient control parameter. In principle,  
cells in tissues could be characterized by a range of adhesion strengths. %, a scenario that we will focus in future studies. 
Competition between these cell types, mediated by adhesion dependent 
pressure feedback into growth, could be critical in determining the relative proportion of cells and therefore the organ size. 
%Thus, the theory explains the simulation results, and by extension the experimental data, nearly quantitatively. 

\appendix
%\begin{footnotesize}
\section{Methods}
\label{appforces}
\begin{figure}
\centering
\includegraphics[width=0.9\textwidth]{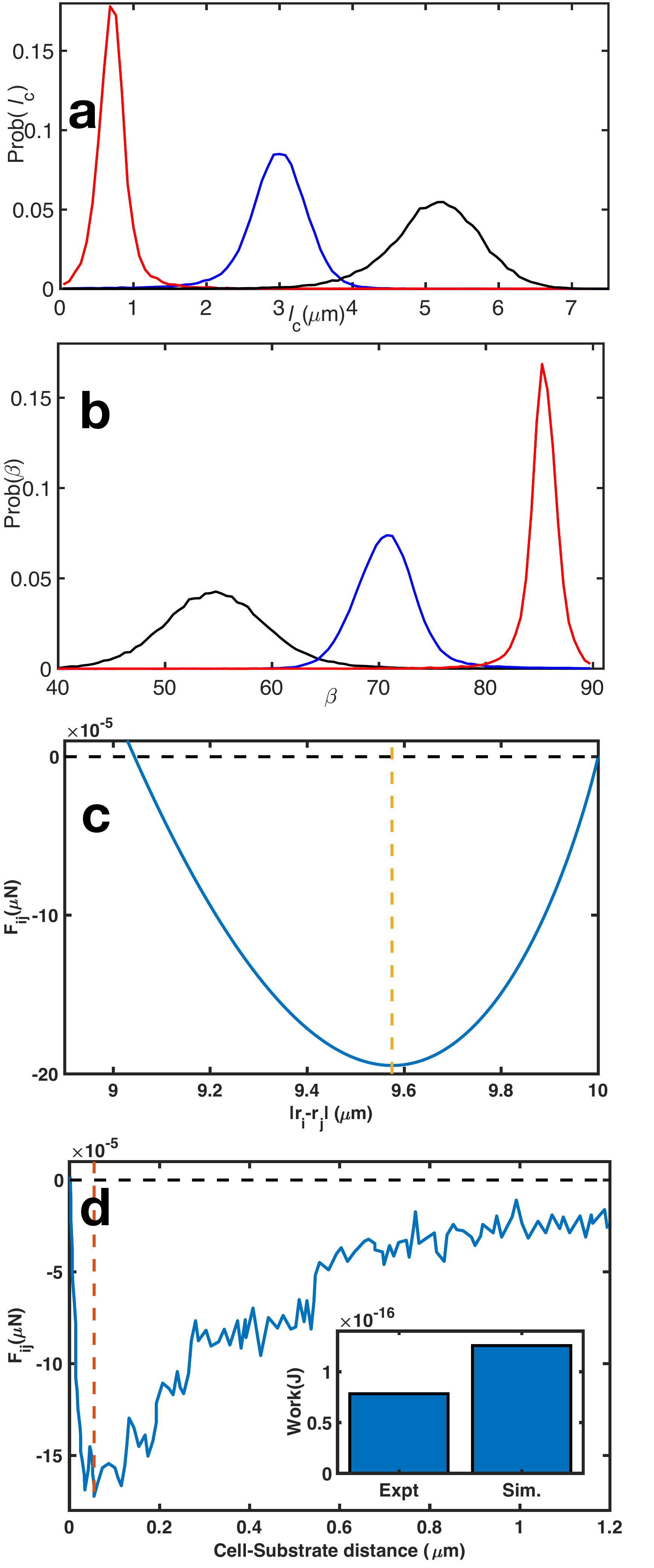}
\caption{{\bf a)} Probability distribution of contact lengths between cells for $f^{ad}=0$ (red), 
$f^{ad}=1.5\times 10^{-4}$ (blue) and $f^{ad}=3\times 10^{-4}$ (black).
{\bf b)} Probability distribution of contact angles, $\beta$ at varying values of $f^{ad}$ with the color scheme as 
in {\bf a)}. {\bf c)} Force on cell $i$ due to $j$, $F_{ij}$, for $R_{i}=R_{j}=5~\mu m$ using mean values of elastic modulus, 
poisson ratio, receptor and ligand concentration (see Table I in the SI). $F_{ij}$ is plotted 
as a function of cell center-to-center distance $|{\mathbf {r}}_{i} -{\mathbf {r}}_{j}|$. 
We used $f^{ad}=1.75\times 10^{-4} \mathrm{\mu N/ \mu m^2}$ to generate $F_{ij}$. {\bf d)} Force-distance data extracted 
from SCFS experiment~\cite{baronsky2016reduction}. Inset shows the work required to separate cell and E-cadherin functionalized substrate 
in SCFS experiment and two cells in theory, respectively. 
Minimum force values are indicated by vertical dashed lines. See Appendix~\ref{appcalibr} for further details.}
\label{forcecomp}
\end{figure}

{\bf Model:} We simulate the collective movement of cells using a minimal 
model of an evolving tumor embedded in a matrix using 
an agent-based three dimensional (3D) model~\cite{schaller2005multicellular,drasdo2005single,galle2005modeling}. 
The cells, which are embedded in a highly viscous material mimicking the extracellular material, 
are represented as deformable objects. 
The inter-cell interactions are characterized by direct elastic (repulsive) and adhesive (attractive) forces. 
The total force on the $i^{th}$ cell is given by, 
\begin{equation}
\vec{F}_{i} = \Sigma_{j \epsilon NN(i)}(F_{ij}^{el}-F_{ij}^{ad})\vec{n}_{ij}, 
\label{netF}
\end{equation}
where $\vec{n}_{ij}$ is the unit vector from the center of cell $j$ to cell $i$. 
The forces are summed over the nearest neighbors ($NN(i)$) of the $i^{th}$ cell. The form of the elastic force, $F^{el}_{ij}$ (see Eq.~\ref{rep}), 
and the inter-cell adhesive force, $F^{ad}_{ij}$ (see Eq.~\ref{ad}), are taken from the study of Schaller and Meyer-Hermann~\cite{schaller2005multicellular}. %, and are 
%described in the Supplementary Information (SI). 
The strength of the adhesive interaction between cells, $f^{ad}$,  
is measured in units of $\mu N/\mu m^{2}$, ($F^{ad}_{ij} \propto f^{ad}$ given by Eq.~\ref{ad}). %ADD EQUATION NUMBER FROM SI
Force as a function of cell center-to-center distance is plotted in Fig.~\ref{forcecomp}c.

Cell-to-cell and cell-to-matrix damping account for the effects of friction 
due to other cells, and the extracellular matrix (ECM) (for example, collagen matrix), respectively. 
%In addition to the systematic forces, the cells  grow stochastically with time, 
%and divide when they reach a critical size. 
The model accounts for apoptosis, cell growth and division. Thus, 
the collective motion of cells is determined by both systematic cell-cell forces and the dynamics 
due to stochastic cell birth and apoptosis under a free boundary condition~\cite{malmi2018cell}.  

{\bf Forces Between Cells and Equations of Motion:} Each cell is represented as a soft sphere whose 
radius changes in time to account for cell growth. 
We characterize each cell by its radius, elastic modulus, 
membrane E-cadherin receptor, and ligand 
concentration. %, characterize each cell. 
Following previous studies~\cite{schaller2005multicellular, drasdo2005single, pathmanathan2009computational}, 
we used Hertzian contact mechanics
to model the magnitude of the elastic force between two spheres of radii $R_{i}$ and $R_{j}$, given by, %(Fig.~\ref{cellcellinter}),
\begin{equation}
\label{rep}
F_{ij}^{el} = \frac{h_{ij}^{3/2}(t)}{\frac{3}{4}(\frac{1-\nu_{i}^2}{E_i} + \frac{1-\nu_{j}^2}{E_j})\sqrt{\frac{1}{R_{i}(t)}+ \frac{1}{R_{j}(t)}}},
\end{equation}
where the parameters $E_{i}$ and $\nu_{i}$, respectively, are the elastic modulus and 
Poisson ratio of the $i^{th}$ cell~\cite{schaller2005multicellular}. The overlap between cells, if they interpenetrate without deformation, 
is $h_{ij}$, defined as $\mathrm{max}[0, R_i + R_j - |\vec{r}_i - \vec{r}_j|]$ with $|\vec{r}_i - \vec{r}_j|$ 
being the center-to-center distance (see Fig.~\ref{angle}b in the Main Text). 
The elastic repulsive forces tend to minimize the overlap between cells, and 
could be thought of as a proxy for cortical tension~\cite{brodland2002differential, hayashi2004surface}. 

 %\par
%The Hertz contact surface area is smaller than the proper spherical contact surface area. However, 
%%in dense tissues many cells overlap, thus the underestimation of the cell surface overlap may be advantageous 
%for realistic values of the adhesion forces~\cite{schaller2005multicellular}. 

The magnitude of the attractive adhesive force, $F_{ij}^{ad}$, between cells $i$ and $j$ is given by, 
\begin{equation}
\label{ad}
F_{ij}^{ad} = A_{ij}f^{ad}\frac{1}{2}(c_{i}^{rec}c_{j}^{lig} + c_{j}^{rec}c_{i}^{lig}),
\end{equation}
where $A_{ij}$ is the cell-cell contact area, $c_{i}^{rec}$ ($c_{i}^{lig}$) is the E-cadherin receptor (ligand) concentration 
(assumed to be normalized with respect to the maximum receptor or ligand concentration such that  
$0 \leq c_{i}^{rec},  c_{i}^{lig} \leq 1$). 
%The receptor and ligand concentration on the cell surface are distributed according to a Gaussian 
%($p(c_{i}^{rec}(c_{i}^{lig}))=\frac{1}{0.02\sqrt{2\pi}} e^{-(c_{i}^{rec}(c_{i}^{lig})-0.9)^2/2\times 0.02^2}$), centered around the mean (=0.9) with a dispersion of $0.02$. 
The coupling constant $f^{ad}$ in Eq.~\ref{ad}, with dimensions $\mu N/ \mu m^2$, allows us to 
%\leq c_{i}^{(rec/lig)/max} 
rescale the adhesion force, to account for the variations in the maximum receptor and ligand concentrations. 
Higher (lower) maximum receptor and ligand concentration on the cell surface membrane, is accounted for by higher (lower) 
value of $f^{ad}$. It should be noted that the strength of adhesion between the cells is mediated by both 
the extracellular portion of E-cadherin and how it interacts with the cytoskeleton. 
The cytoplasmic E-cadherin domain, in conjunction with $\alpha$-catenin, 
binds to $\beta$-catenin, linking it to the actin cytoskeleton~\cite{hayashi2004surface}. 
In the minimal model, all of these complicated processes that occur
on sub-cellular length scales are subsumed in $f^{ad}$. % the cell-cell adhesion strength. %with dimensions $\mu N/ \mu m^2$ implicitly assumed. 
The inter cell contact surface area, $A_{ij}$ (see Eq.~\ref{ad}), is obtained using 
the Hertz model prediction,  $A_{ij} = \pi h_{ij}R_{i}R_{j}/(R_{i}+R_{j})$~\cite{schaller2005multicellular}. 

\begin{figure}
\centering
\includegraphics[width=\textwidth]{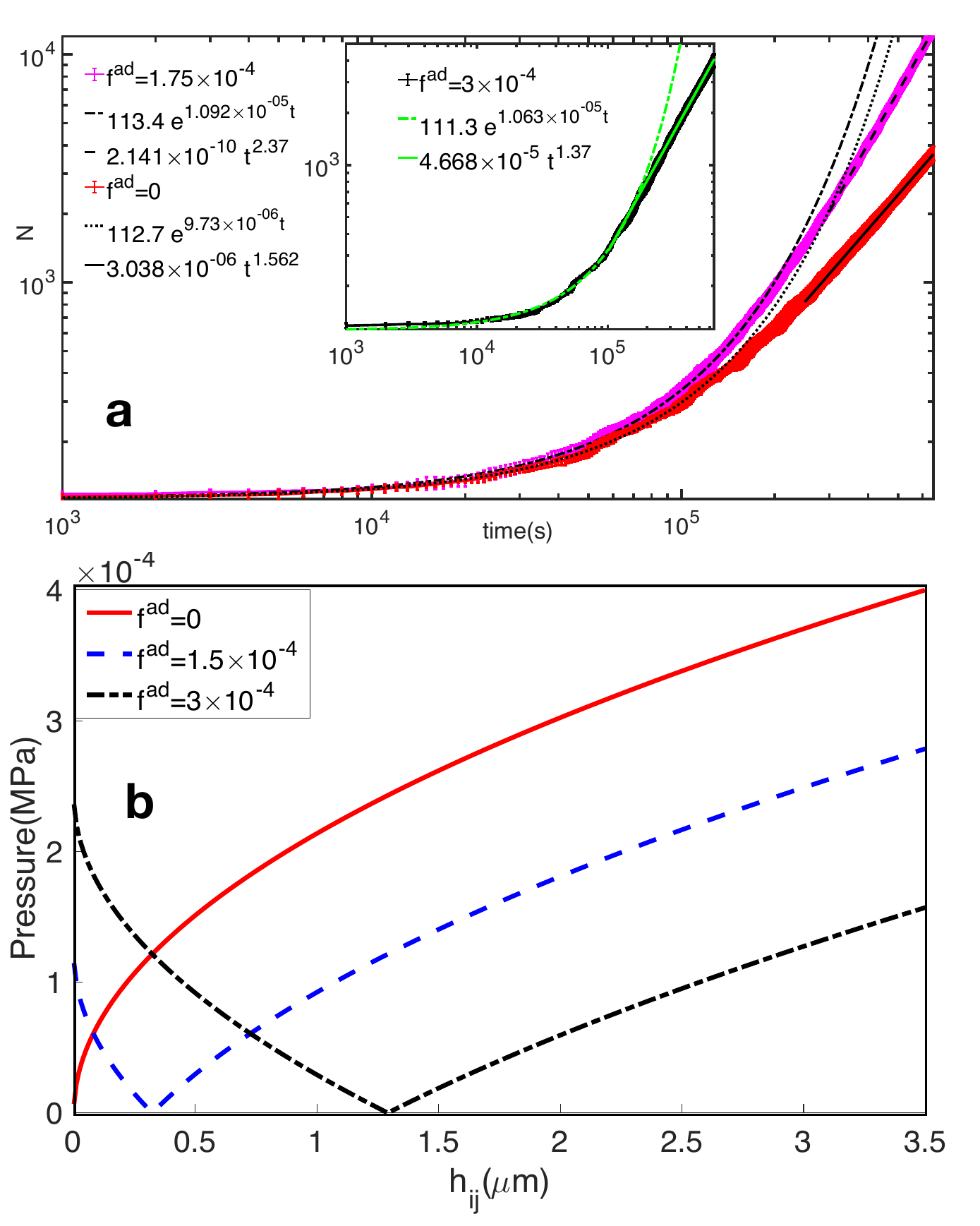}
\caption{{\bf a)} Number of cells, $N(t)$, over 7.5 days of growth. Initial exponential growth 
followed by power law growth behavior is seen for three different $f^{ad}$ values. 
The onset of power-law growth in $N$ occurs between $t=10^{5} - 2\times 10^{5}$ secs. 
Power law exponent depends on $f^{ad}$. 
{\bf b)} Pressure, $p_i$, experienced by a 
cell interacting with another cell as a function of overlap distance ($h_{ij}$) for different values of $f^{ad}$. 
$F_{ij}$ is calculated for $R_{i}=R_{j}=4~\mu m$ using mean values of elastic modulus, 
poisson ratio, receptor and ligand concentration (see Table I).}
\label{p1ress}
\end{figure}

\begin{figure}
\centering
\includegraphics[width=\textwidth]{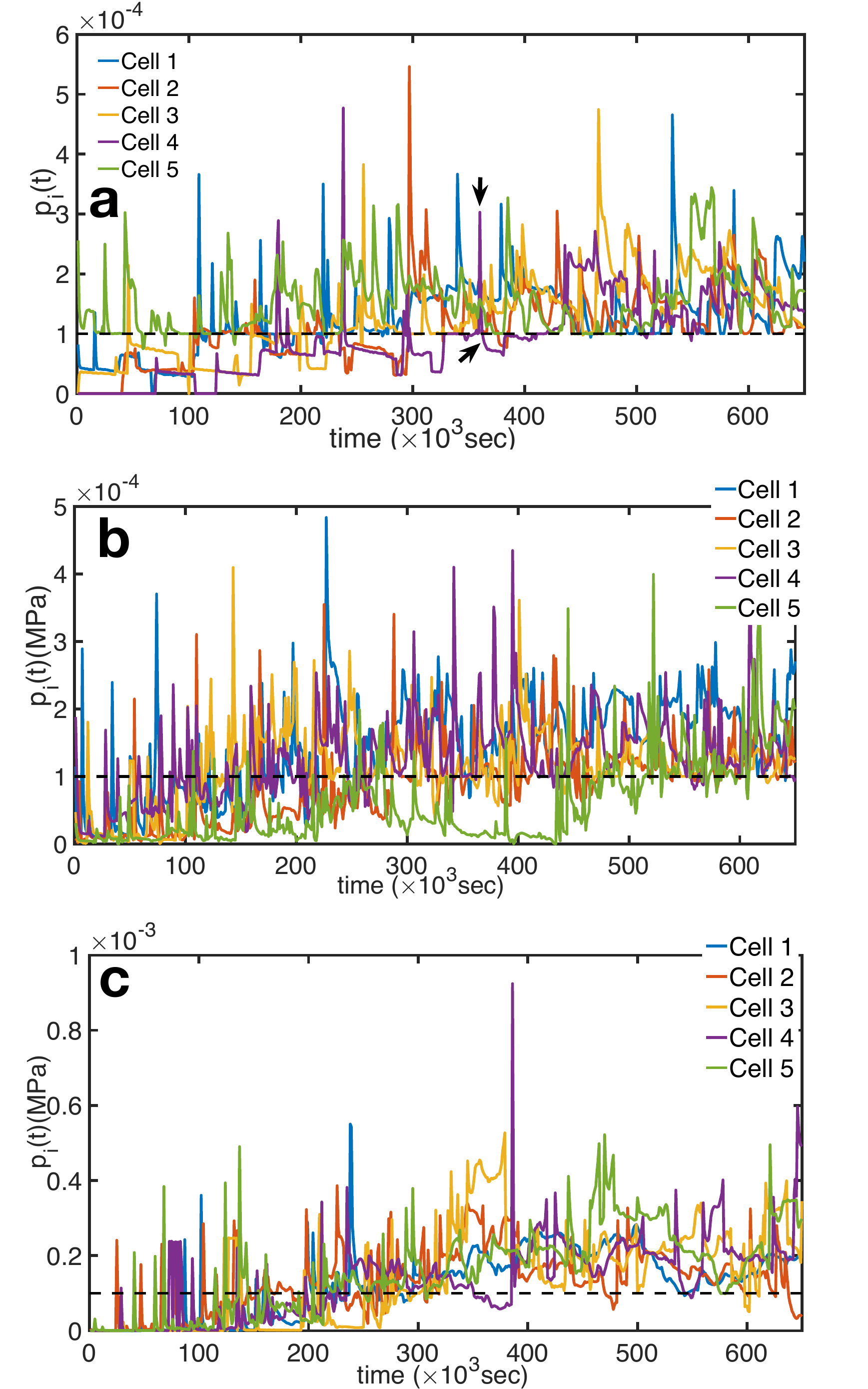}
\caption{{\bf a)} Pressure experienced by individual cells as a function of time at $f^{ad}=0$,
{\bf b)} $f^{ad}=1.5\times 10^{-4}$ and {\bf c)} $f^{ad}=3\times 10^{-4}$. Dashed black lines 
indicate the critical pressure, $p_c$. Black arrows in {\bf a)} highlight fluctuations 
in $p_i$ above and below $p_c$.}
\label{press}
\end{figure}

An optimal range of cell-cell 
packing ($h_{ij}$) exists where $p_i$ is minimized (see Fig.~\ref{p1ress}b). 
With the definition $p_{i}$, 
growth in the total number of cells ($N(t)$) is well approximated 
as an exponential $N(t) \propto {\mathrm exp(const\times t)}$ 
at short times ($t<10^{5}~\mathrm{secs}$) (see Fig.~\ref{p1ress}a). %as showed by the exponential fitting of solid red line to the simulation results (purple squares). 
At longer time scales ($t>\sim 3\times 10^{5}~\mathrm{secs}$), the increase in 
the tumor size follows a power law, $N(t) \propto t^{\beta}$. 
%quickly as we reduce the $p_{c}$ although it 
%while $N(t)$ retains exponential growth at early stages
Such a cross over from exponential to power-law growth in 3D tumor spheroid size has been 
observed in many tumor cell lines with $\beta$ varying 
from one to three~\cite{conger,Mandonnet2003,Simeoni2004,Hart1998,grimes}. 
Based on the power law growth behavior, we expect, $\Delta N(f^{ad},t)/N = [N(f^{ad}_{c}=1.75 \times 10^{-4} \mu N/\mu m^2,t) - N(f^{ad},t)]/N(f^{ad}_{c}=1.75 \times 10^{-4} \mu N/\mu m^2,t)$, 
for $f^{ad} = 0$ to be $1-(3\times 10^{-6} t^{1.6})/(2.1\times 10^{-10} t^{2.4})$. Hence, $\Delta N(f^{ad}=0,t)/N=1-A_{0}t^{-0.8}$, where $A_{0}= 1.4\times 10^{4}$ is a constant. 
Similarly for $\Delta N(f^{ad}=3\times 10^{-4}  \mu N/\mu m^2,t)/N= 1-A_{1}t^{-1}$ where $A_{1}=2.2\times 10^{5}$.  Due to the local cell density
fluctuations caused by the birth-death events, the cell pressure $p_{i}$ is a highly dynamic quantity (see Figs.~\ref{press}a-c). $p_{i}(t)$ 
plays an important role in how local forces provide a feedback on cell growth and division. 

%In the model, 
%higher (lower) value of $f^{ad}$, is a proxy for higher (lower) cell adhesion molecule (CAM) expression 
%levels on the cell membrane. 
While cells are characterized by many different types of cell adhesion molecules (CAMs), 
here we focus on E-cadherin. 
We consider different levels of CAM expression, varying from low ($f^{ad}=0~\mathrm{\mu N/\mu m^2}$) to 
intermediate ($f^{ad}=1.5 ~\mathrm{\mu N/\mu m^2}$) to high ($f^{ad}=3~\mathrm{\mu N/\mu m^2}$) values. 
For a discussion on the appropriateness of the value of the range of $f^{ad}$, see Appendix \ref{appcalibr}. 
%Repulsive and adhesive forces, given in Eqs.~(\ref{rep}) and (\ref{ad}), 
%%act on the center of the spheres 
%act along the unit vector $\vec{n}_{ij}$ pointing from the centers of cells $j$ to $i$.
%The total force on the $i^{th}$ cell is given by the sum of forces over its nearest neighbors ($NN(i)$), 
%\begin{equation}
%\vec{F}_{i} = \Sigma_{j \epsilon NN(i)}(F_{ij}^{el}-F_{ij}^{ad})\vec{n}_{ij}. 
%\label{netF}
%\end{equation}
%where %\begin{equation}
%%\label{rep}
%$F_{ij}^{el} = \frac{h_{ij}^{3/2}(t)}{\frac{3}{4}(\frac{1-\nu_{i}^2}{E_i} + \frac{1-\nu_{j}^2}{E_j})\sqrt{\frac{1}{R_{i}(t)}+ \frac{1}{R_{j}(t)}}}$
%is the elastic repulsive force. 
%\end{equation}
%Let us call $F_{ij}=F_{ij}^{el}-F_{ij}^{ad}$, thus 
We used a distance sorting algorithm to efficiently obtain a list of nearest neighbors in contact with the $i^{th}$ cell.  
For a cell, $i$, an array with distances from cell $i$ to all the other cells 
is created. We then calculated $R_i + R_j - |\vec{r}_i - \vec{r}_j|$ and sorted for cells $j$
that satisfy the condition $R_i + R_j - |\vec{r}_i - \vec{r}_j|~>~0$, a necessary condition for any cell $j$ to be in contact with cell $i$. 

The justification that the inertial forces can be neglected
can be found in our previous study (see Fig. 18 in Ref.~\cite{malmi2018cell}). 
If we neglect inertial effects, the equation of motion of the $i^{th}$ cell is, 
\begin{equation}
\label{eqforce}
\dot{\vec{r}}_{i} = \frac{\vec{F}_{i}}{\gamma_i}, 
\end{equation}
where, $\gamma_i = \gamma_{i}^{\alpha' \beta', visc} + \gamma_{i}^{\alpha' \beta', ad} $ is the friction coefficient 
with 
\begin{equation}
\gamma_{i}^{\alpha' \beta', visc} = 6\pi \eta R_{i} \delta^{\alpha' \beta'},
\end{equation}
and 
\begin{eqnarray}
\gamma_{i}^{\alpha' \beta', ad} =&& \gamma^{max}\Sigma_{j \epsilon NN(i)} (A_{ij}\frac{1}{2}(1+\frac{\vec{F}_{i} \cdot \vec{n}_{ij}}{|\vec{F}_{i}|})\times \\ \nonumber &&\frac{1}{2}(c_{i}^{rec}c_{j}^{lig} + c_{j}^{rec}c_{i}^{lig}))\delta^{\alpha' \beta'} \, ,
\end{eqnarray}
being the cell-to-matrix and cell-to-cell damping contributions respectively. Here, the indices $\alpha'$, $\beta'$ represent 
cartesian co-ordinates. 
Viscosity of the medium surrounding the cell is denoted by $\eta$ and $\gamma^{max}$ is the adhesive 
friction coefficient. Additional details of the simulation methods are given elsewhere~\cite{malmi2018cell}. 
Note that because the equations of motion for the coarse-grained model contain the friction term they do not satisfy Galilean invariance. 

{\bf Pressure-dependent Dormancy:} A crucial feature in the model is the role 
played by the local pressure, $p_i$, experienced by the $i^{th}$ cell relative to a critical pressure, $p_{c}$. 
A given cell, at any time $t$, can either be in the 
dormant ($D$) or in the growth ($G$) phase depending on the pressure on the cell (see Fig~\ref{angle}b). 
The total pressure ($p_{i}$) on the $i^{th}$ cell,
\begin{equation}
\label{pre}
p_i = \sum_{j \in NN(i)} \frac{|{F_{ij}}|}{A_{ij}}, 
\end{equation} 
%The pressure experienced by the cell, $p_{i}$, 
is the overall sum of all the normal pressures due to the nearest neighbors. 
%We also considered the Irving-Kirkwood definition of local pressure, and found that our results 
%are robust under this alternative definition of pressure as we discuss in detail later. %(see SI Fig. S4). 
If $p_{i} > p_c$ (a pre-assigned value of the critical pressure), the cell 
becomes dormant, and can no longer grow or divide.  %add SI figure
Note that a cell that becomes dormant at time $t$ does not imply 
that it remains so at all later times because as the cell colony evolves, $p_i$, a dynamic 
quantity fluctuates, and hence can become less or greater than $p_c$ (see Figs.~\ref{press}a-c, Supplementary Movies 1-3A). 
It has been shown {\it in vitro} that solid stress, defined as the mechanical stress due to solid and elastic 
elements of the extracellular matrix, inhibits growth of multicellular tumor spheroids 
irrespective of the host species, tissue origin or differentiation state~\cite{helmlinger1997solid}. 
This type of growth inhibition is mediated by 
stress accumulation around the spheroid as a result 
of the progressive displacement of the surrounding matrix due 
to the growing clump of cells. 
In our model, however, 
the effect of pressure on cell growth is driven by local cell-cell contact as opposed to 
the global stress exerted by the surrounding matrix. 
Both {\it in vivo} and {\it in vitro}, epithelial cells exhibit 
contact inhibition of proliferation due to cell-cell interactions~\cite{eagle1967growth}. %and cell-matrix 
The value of the critical pressure used in our work is in the same range as experimentally measured  
cell-scale stresses (10-200 Pa)~\cite{mongera2018fluid} and with 
earlier works using critical cellular compression as the mechanism for contact 
inhibition~\cite{schaller2005multicellular,galle2005modeling}. 
We have also verified that the qualitative results are independent of the precise value of 
$p_c$, as well as alternative definitions of local pressure (see Sections I-II in the SI). 

{\bf Cell Dynamics:} Because the Reynolds number for cells in a tissue is 
small~\cite{schaller2005multicellular,dallon}, overdamped 
approximation is appropriate. The equations of motion are given below (see Eq.~\ref{eqforce}). 
Besides the cell-cell repulsive and adhesive forces, another contribution to cell dynamics 
comes from cell growth, division and apoptosis. Stochastic cell growth leads to dynamic variations 
in the cell-cell forces. Cell division and apoptosis induce temporal rearrangements in the cell positions 
and packing. Hence, both the contribution of the systematic forces and cell growth, birth and apoptosis 
towards cell dynamics are taken into account in the model, for which we described the unusual 
dynamics previously~\cite{malmi2018cell,samanta2019origin}. 
The parameters used in the simulations are given in the SI (see Table I). We   
justify the range of $f^{ad}$ explored in our simulations in Appendix~\ref{appcalibr}. 
In order to explore the plausible dual role of E-cadherin on 
cell proliferation, we vary $f^{ad}$, keeping all other parameters constant. 
%\end{footnotesize}

\section{Calibration of Cell-Cell Adhesion Strength}
\label{appcalibr}
The crucial parameter in the present study is the cell-cell interaction 
strength, $f^{ad}$, which is a proxy for E-cadherin expression. In order to assess if the values 
used in our simulations are in a reasonable range, we estimated $f^{ad}$ from the typical strength of cell-cell attractive 
interactions reported in previous studies. Early experiments showed that the interaction strength between cell adhesion 
proteoglycans is $\sim 2\times 10^{-5} \mu N/\mu m^{2}$~~\cite{Dammer95Science}. 
More recently, single cell force spectroscopy (SCFS) technique has been used to measure directly the typical forces
required to rupture E-cadherin mediated bonds between cells. Several types of cadherins 
could be present on the cell surface, in addition to  
adhesion molecules such as integrins, selectins etc~\cite{van2008cell}.
In order to confirm that it is indeed E-Cadherin expression level that changes  
at different stages of embryo development, Baronsky et. al~\cite{baronsky2016reduction} 
functionalized gold coated substrate with E-cadherin, and measured the force-distance curves between 
primordial germ cells and the substrate. 

\begin{figure}
\centering
\includegraphics[width=0.95\textwidth]{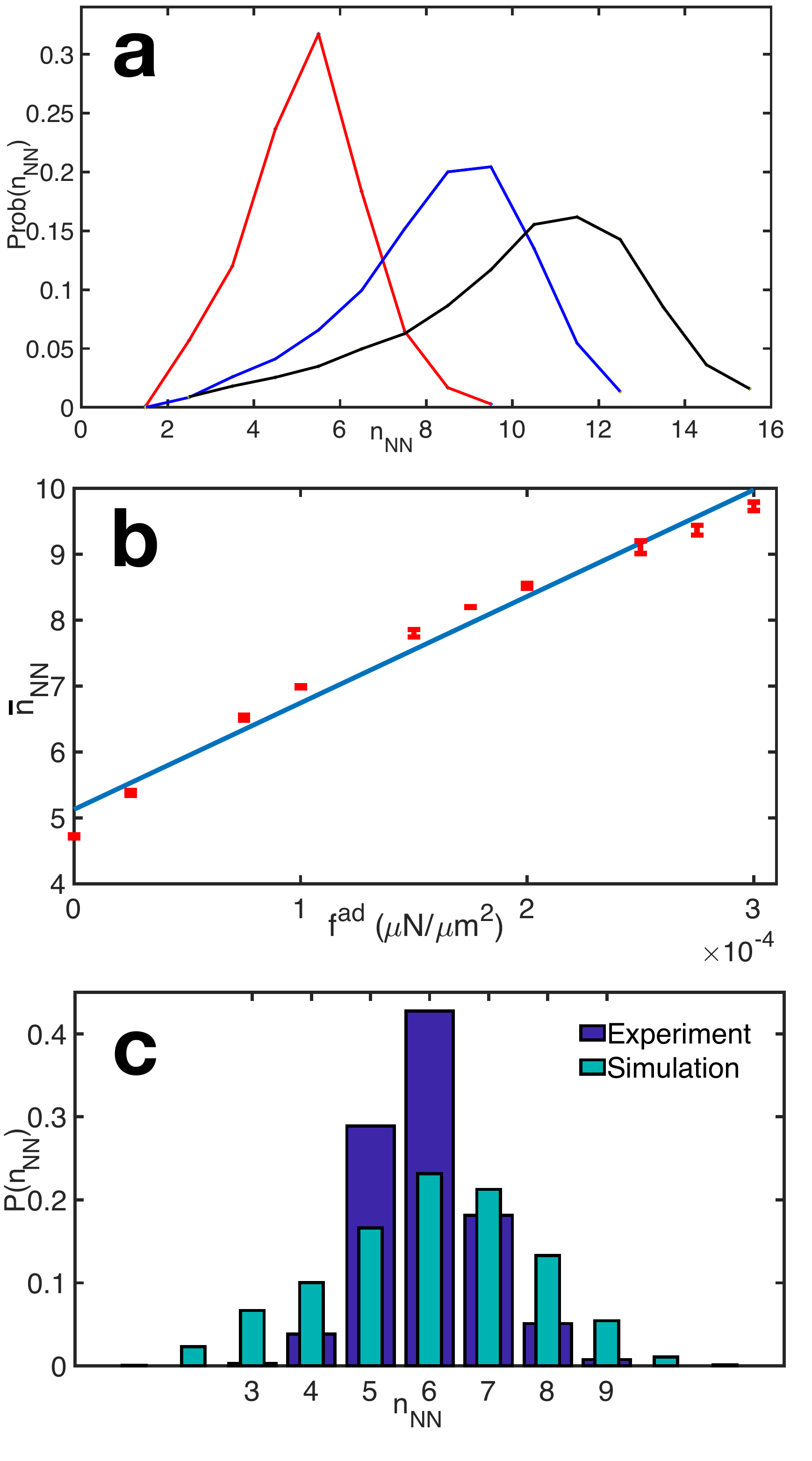}
\caption{{\bf a)} Graph showing the probability 
distribution of the number of nearest 
neighbors  ($n_{NN}$) on day 7.5 of tumor growth. Red, blue and black curves 
are for $f^{ad}=0, 1.5\times 10^{-4} \mathrm{\mu N/\mu m^{2}},~\mathrm{and} ~3\times 10^{-4} \mathrm{\mu N/\mu m^{2}},$ respectively. 
{\bf b)}Average number of nearest 
neighbors, $\bar{n}_{NN}$, as a function of $f^{ad}$. 
Linear fit shows $\bar{n}_{NN} \sim 5.1 + (1.6 \times 10^4)\times f^{ad}$ ($G+\beta \times f^{ad}$). 
Error bars represent the standard deviation. 
{\bf c)} Comparison of the probability distribution of the 
number of nearest neighbors  ($n_{NN}$) between experiments and simulations. 
Experimental data is for Xenopus tail 
epidermis (n=1,051 cells)~\cite{gibson2006emergence}. 
Simulation data is for  
$f^{ad}=5\times 10^{-5}~\mathrm{\mu N/\mu m^2}$. Data in (a) - (c) are from 3 independent simulation runs. }
\label{figsi2}
\end{figure}

%In our simulations, the cell-cell adhesion strength is modulated by the parameter $f^{ad}$.
Within the range of $f^{ad}$ considered in the simulations, a typical force distance curve (plot of $|\vec{F}_{i}|$ versus 
$|\vec{r}_{i}-\vec{r}_{j}|$ from Eq.~\ref{netF}) is shown in Fig.~\ref{forcecomp}c. 
The plot in Fig.~\ref{forcecomp}c shows that for typical cell sizes ($\approx 5 \mu m$) the minimum force is $\approx 2\times 10^{-4} \mu N$, 
which is fairly close to the values $\approx 1.5\times 10^{-4} \mu N$ in Fig.~\ref{forcecomp}d and $4\times 10^{-4} \mu N$ 
reported elsewhere~\cite{Dammer95Science}.  The inset in Fig.~\ref{forcecomp}d shows the work done to overcome the adhesion 
force mediated by E-cadherin which is the area under the force-distance curve (FDC). 
 The work expended is comparable at $0.78\times 10^{-16} \mathrm{Nm}$ for primordial germ cells (PGCs in {\it Xenopus laevis} embryo) to E-cadherin substrate 
separation experiment and $1.25\times 10^{-16} \mathrm{Nm}$ for cell-cell separation in the model. 
Because the set up in single cell force spectroscopy and the theoretical 
model are not precisely comparable, it is gratifying that the magnitude of forces required 
to separate two cells obtained using Eq.~(\ref{netF}) and the measured values are not significantly different. Note that we did not
adjust any parameters to obtain the reasonable agreement.
%We point this out to merely justify the choice of the parameters used in our simulations. 
We undertook this comparison to merely point out that the range of 
E-cadherin mediated forces used in our simulations 
reflects the typical cell-cell adhesion strength measured in experiments. %\par

The timescale associated with single receptor-ligand binding is typically 
2-10 seconds~\cite{galle2005modeling}. 
There are about $\sim 10$ cadherins/$\mu m^{2}$ on the surface of typical cells~\cite{amack2012knowing},
corresponding to $\sim$3000 cadherins on the 
cell surface for a cell with radius of 5 microns. For studies of cell growth and dynamics on the 
time scale of days, the fluctuations at the level of single receptor-ligand binding can therefore be 
neglected~\cite{galle2005modeling}, thus justifying the use of constant $f^{ad}$ values. The receptor/ligand concentration 
are sampled from a Gaussian distribution (see SI Section IV for more details). 

Given the center-to-center distance, $r_{ij}=|r_{i}-r_{j}|$, between cells $i$ and $j$, the contact 
length, ${\it l}_{c}$, and contact angle $\beta$ can be calculated. Let $x$ be the distance from center of cell $i$ to 
contact zone marked by ${\it l}_{c}$, along $r_{ij}$. Similarly, we define $y$ as the distance between 
center of cell $j$ to ${\it l}_{c}$ once again along $r_{ij}$ (see Fig.~\ref{angle}a of Main Text). Based on the right triangle 
that is formed between $x$, $R_i$ and ${\it l}_{c}/2$, $R_{i}^{2}-x^{2} = R_{j}^{2}-y^{2}=({\it l}_{c}/2)^2$ and 
$x+y=r_{ij}$. This allows us to solve for $x,y$ and hence, 
\begin{eqnarray}
{\it l}_{c} =&& 2\sqrt{\frac{4r_{ij}^{2}R_{i}^{2} -(r_{ij}^{2}+R_{i}^{2}-R_{j}^{2})^{2}}{4r_{ij}^{2}}}, \\
\beta =&& arctan(2y/ {\it l}_{c}).
\end{eqnarray}
The probability distribution for ${\it l}_{c}$ and $\beta$ obtained from the simulation for 
varying values of $f^{ad}$ is shown in Figs.~\ref{forcecomp}a - \ref{forcecomp}b.

The interaction parameters characterizing the model are, the two elastic constants ($E_{i}$ and $\nu_{i}$) and 
$f^{ad}$ if we assume that the combination of receptor and ligand concentrations in Eq.~(\ref{ad}) is a constant. 
In addition, the evolution of cell colony introduces two other parameters, birth ($k_{b}$) and apoptotic rates ($k_{a}$) 
of cells. The values of $k_{a}$ and $k_{b}$ depend on the detailed biology governing cell fate, which we
simply take as parameters in the simulations. If the elastic constants, and $k_{a}$, $k_{b}$ are fixed, then the only parameter 
that determines the evolution of the tumor is $f^{ad}$ and $p_{c}$, whose magnitude is determined by E-cadherin 
expression. Here, we explore the effects of $f^{ad}$ and $p_{c}$, %which is determined by the extent of E-cadherin expression, 
on tumor proliferation. 

\section{Average Number of Nearest Neighbor of Cells Increases with \texorpdfstring{$f^{ad}$}{Lg}}
 \label{appaverage}
%Understanding the effect of cell-cell adhesion on growth and 
%proliferation of cells is of immense importance in the 
%context of morphogenesis and tumorigenesis. 
%To calibrate the cell-cell adhesion levels used in 
%the simulation, we study how 
The collective movement of cells (related to proliferative capacity) is determined 
by cell arrangement and packing within 
the three dimensional (3D) tissue, which clearly depends on the adhesion strength. 
Analyzing the distribution of number of nearest neighbors (see Fig.~\ref{figsi2}a), the arrangement of cells in 
the spheroid has a peak near $6$ nearest neighbors at low $f^{ad}$. 
Very few cells, if any, have less than $2$ neighbors. 
Similarly, at %the low adhesion strength of 
$f^{ad}=5\times 10^{-5}~\mathrm{\mu N/\mu m^2}$, few cells have 
more than $9$ neighbors (see Fig.~\ref{figsi2}c). 
As $f^{ad}$ increases, 
the peak in the nearest neighbor distribution 
moves to higher values, with the 
distribution also becoming broader (Fig.~\ref{figsi2}a). 
With the highest adhesion strength, $f^{ad}=3\times 10^{-4} \mathrm{\mu N/\mu m^2}$, 
the average number of nearest neighbors is $\approx 9$ cells which is consistent 
with 3D experimental data for mouse blastocyst after $5-9$ days of growth~\cite{fischer2017three}.

We surmise that the dependence of the average number of nearest neighbors on cell-cell adhesion strengths $f^{ad}$ in the simulations  
is consistent with experimental findings. Cell packing data in 2D epithelial 
structures, quantified by the probability distribution of nearest neighbors~\cite{gibson2006emergence},   
allow us to compare the simulation results to experiments (Fig.~\ref{figsi2}c). 
%for the range  of $f^{ad}$ considered, 
We compare the simulation results for the distribution of the number of nearest neighbors ($n_{NN}$) with experiment, 
keeping in mind that our simulation is in 3D. 
In 3D, the average number of nearest neighbors is 
higher than in 2D. %due to the additional spatial dimension. 
There is indeed an increase in the probability of nearest neighbors 
from $7-10$ (Fig.~\ref{figsi2}c).  
Moreover, $f^{ad} >100~\mathrm{dynes/cm^2} = 10^{-5} \mathrm{\mu N/\mu m^2}$, within an order of magnitude  
has been reported in experiments~\cite{byers1995role}, which we point out only to show that the values of $f^{ad}$ 
considered in our study are reasonable. \par
%For cell to substrate adhesion, adhesion strengths of $f_{cell-substrate}^{ad} \sim  10^{-5} - 7.5\times 10^{-5} \mathrm{\mu N/\mu m^2}$ has been 
%reported~\cite{singh2013adhesion}.

The average number of nearest neighbors, $\bar{n}_{NN}$, increases as $f^{ad}$ increases 
(Fig.~\ref{figsi2}b). %leads to the conclusion that a cell is in contact with more neighboring cells as $f^{ad}$ increases. 
The approximate linear fit %to the adhesion strength dependence of the average $n_{NN}$ is obtained 
$\bar{n}_{NN} \sim G+\beta f^{ad}$ is used to rationalize the data in Fig.~\ref{nvst}a in the Main Text. 
The fit parameters $G,\beta$ are given in the caption. The linear fit is used only for calculating 
$f^{ad}_{c}$, the optimal value at which proliferation is a maximum. 

\begin{figure}
\centering
\includegraphics[width=1\textwidth]{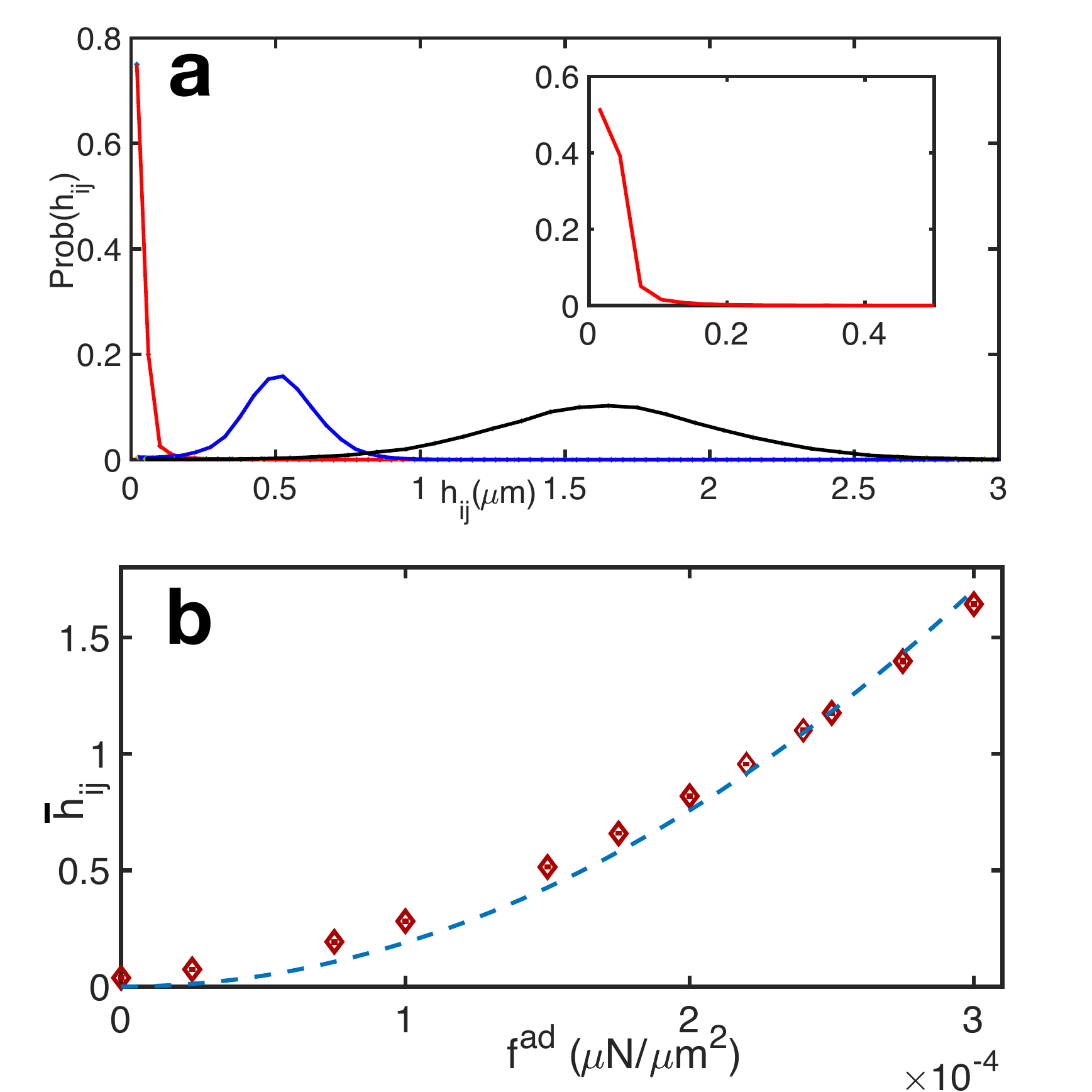}
%\sidesubfloat[]{\includegraphics[width=0.49\textwidth] {prbhij_3runavg_curves.eps} \label{figs1a}}
%\sidesubfloat[]{\includegraphics[width=0.49\textwidth] {hij_avg_3runavg.eps} \label{figs1b}}
\caption{{\bf a)} Probability distribution of the
overlap ($h_{ij}$) of cells on day 7.5 of tumor 
growth for three different adhesion strengths. Red, blue and black curves 
are for $f^{ad}=0, 1.5\times 10^{-4} \mu N/\mu m^{2}, \mathrm{and}  ~3\times 10^{-4} \mu N/\mu m^{2},$ respectively. 
{\bf b)} Average interpenetration distance has a 
quadratic dependence on adhesion strength - $\bar{h}_{ij} \sim (4350 f^{ad})^2$. 
Data obtained are from 3 independent 
simulation runs. Error bars represent the standard deviation. }
\label{figsi1}
\end{figure}

{\bf Distribution of h$_{ij}$:} The cell-cell overlap, $h_{ij}$, gives 
an indication of how closely the deformable cells are packed within the 3D spheroid. 
At low adhesion strengths, the distribution 
is sharply peaked at small $h_{ij}$ (see Fig.~\ref{figsi1}a), implying there is minimal cell-cell interpenetration. 
As $f^{ad}$ increases, the cells are jammed.  
For $f^{ad}=3\times 10^{-4}~\mathrm{\mu N/\mu m^2}$, the average cell overlap $\bar{h}_{ij} \approx 1.6~\mathrm{\mu m}$, 
implying that the center to center distance between 
cells is approximately $6.4~\mathrm{\mu m}$ (for cells of radii $4~\mathrm{\mu m}$). 
Note that the cell overlap distribution becomes broader 
as $f^{ad}$ increases. The mean overlap, 
$\bar{h}_{ij}$, varies quadratically with  
adhesion strength (Fig.~\ref{figsi1}b). 
If we set $F^{el} = F^{ad}$, we find that $h\sim(f^{ad})^2$, and as 
expected we obtain the fit $\bar{h}_{ij} \approx K(f^{ad})^2$. \par

\begin{figure}
%\begin{turn}{-90}
\centering
\includegraphics[width=\textwidth] {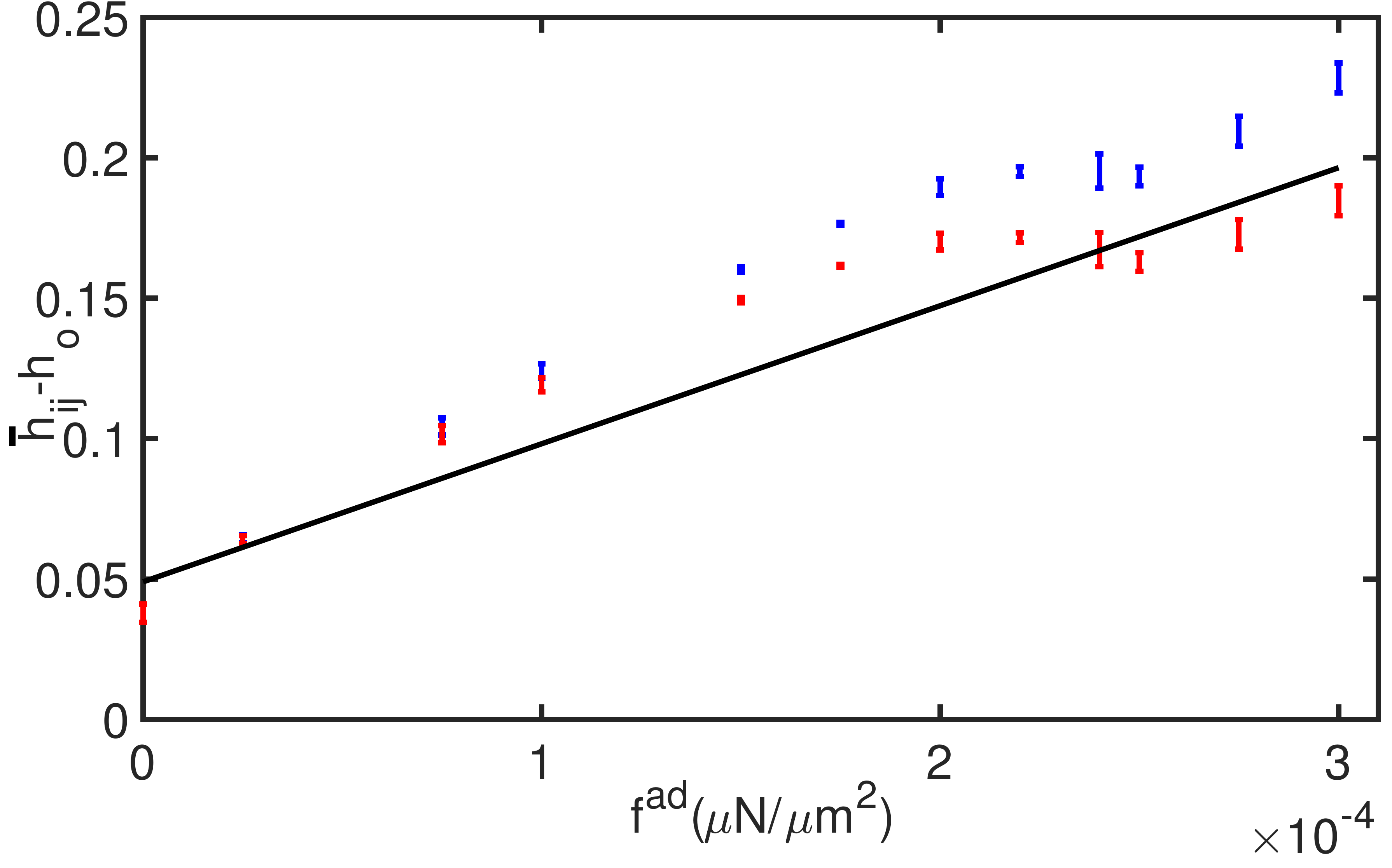} %oa
%\end{center}
%\end{turn}
\caption{Deviation of the average $h_{ij}$ 
($\bar{h}_{ij}$) 
from $h_0$ (cell interpenetration distance 
with minimum possible pressure) for differing adhesion strength showing 
approximate linear dependence $\bar{h}_{ij} -  h_0 \sim 0.05+(491.7)\times f^{ad}$ ($E+\alpha \times f^{ad}$). The value of $h_0$ is determined by  
the radii of any two interacting cells. Two sets of points are shown, blue bars 
are for average radii of $R_{i}=R_{j}=3.9\mu m$ and the 
reds are for $R_{i}=R_{j}=4\mu m$.} %inset pressure figure showing h_0 would be appropriate here.
\label{figsi3}
\end{figure}

To calculate the total pressure 
experienced by a cell ($p_{t}$) theoretically, as detailed in the Main text, 
we look at the deviation of the average cell overlap $\bar{h}_{ij}$ %(or equivalently $\langle h_{ij} \rangle$) 
from the optimal overlap ($h_0$) as $f^{ad}$ is changed, where 
$h_0$ is the overlap distance 
at which the pressure experienced by a cell 
is a minimum. This would occur when 
the repulsive and attractive interaction 
forces between a pair of cells balance ($F^{el} = F^{ad}$).
The deviation, $\bar{h}_{ij} -  h_0$, increases as $f^{ad}$ increases,  
an indication that it is harder for cells to relax to optimal intercellular distances, 
%at higher adhesion strengths. 
due to packing frustration. 
For the purposes of rationalizing the optimal value of $f^{ad}$ (see Main Text) we write, %$E, \alpha$ identified 
$\bar{h}_{ij} -  h_0 = E+\alpha f^{ad}$, where the parameters $E, \alpha$ are as listed in Fig.~\ref{figsi3}. 
We found that $h_0$ depends on radii of the cells that are in contact as well as other parameters, 
\begin{equation}
h_{0} = 3.6 (\frac{1-\nu_{i}^2}{E_i} + \frac{1-\nu_{j}^2}{E_j})^{2} (\frac{R_{i}R_{j}}{R_{i}+R_{j}})(f^{ad})^{2}.
\end{equation}
Hence, we estimate $\bar{h}_{ij} -  h_0$ by using average quantities. 

%\clearpage
{\bf Acknowledgements:} We acknowledge Anne D. Bowen at the Visualization
Laboratory (Vislab), Texas Advanced Computing Center,
for help with video visualizations. This work is supported by the National Science Foundation (PHY 17-08128 and CHE 16-32756) 
and the Collie-Welch Chair through the Welch Foundation (F-0019).

%\showacknow

\clearpage
\bibliography{main2} 

\begin{thebibliography}{10}

\bibitem{legoff2016mechanical}
Lo{\"\i}c LeGoff and Thomas Lecuit.
\newblock Mechanical forces and growth in animal tissues.
\newblock {\em Cold Spring Harbor perspectives in biology}, 8(3):a019232, 2016.

\bibitem{friedl2017tuning}
Peter Friedl and Roberto Mayor.
\newblock Tuning collective cell migration by cell--cell junction regulation.
\newblock {\em Cold Spring Harbor perspectives in biology}, 9(4):a029199, 2017.

\bibitem{budnar2013mechanobiological}
Srikanth Budnar and Alpha~S Yap.
\newblock A mechanobiological perspective on cadherins and the actin-myosin
  cytoskeleton.
\newblock {\em F1000prime reports}, 5(35):1--6, 2013.

\bibitem{tambe2011collective}
Dhananjay~T Tambe, C~Corey Hardin, Thomas~E Angelini, Kavitha Rajendran,
  Chan~Young Park, Xavier Serra-Picamal, Enhua~H Zhou, Muhammad~H Zaman,
  James~P Butler, David~A Weitz, et~al.
\newblock Collective cell guidance by cooperative intercellular forces.
\newblock {\em Nature materials}, 10(6):469, 2011.

\bibitem{halbleib2006cadherins}
Jennifer~M Halbleib and W~James Nelson.
\newblock Cadherins in development: cell adhesion, sorting, and tissue
  morphogenesis.
\newblock {\em Genes \& development}, 20(23):3199--3214, 2006.

\bibitem{takeichi1988cadherins}
Masatoshi Takeichi.
\newblock The cadherins: cell-cell adhesion molecules controlling animal
  morphogenesis.
\newblock {\em Development}, 102(4):639--655, 1988.

\bibitem{saito2012classical}
Masataka Saito, Dana~K Tucker, Drew Kohlhorst, Carien~M Niessen, and Andrew~P
  Kowalczyk.
\newblock Classical and desmosomal cadherins at a glance.
\newblock {\em J Cell Sci}, 125(11):2547--2552, 2012.

\bibitem{tabdanov2009role}
Erdem Tabdanov, Nicolas Borghi, Fran{\c{c}}oise Brochard-Wyart, Sylvie Dufour,
  and Jean-Paul Thiery.
\newblock Role of e-cadherin in membrane-cortex interaction probed by nanotube
  extrusion.
\newblock {\em Biophysical journal}, 96(6):2457--2465, 2009.

\bibitem{borghi2012cadherin}
Nicolas Borghi, Maria Sorokina, Olga~G Shcherbakova, William~I Weis, Beth~L
  Pruitt, W~James Nelson, and Alexander~R Dunn.
\newblock E-cadherin is under constitutive actomyosin-generated tension that is
  increased at cell--cell contacts upon externally applied stretch.
\newblock {\em Proceedings of the National Academy of Sciences},
  109(31):12568--12573, 2012.

\bibitem{thompson1942growth}
Darcy~Wentworth Thompson et~al.
\newblock On growth and form.
\newblock {\em Cambridge Univ. Press}, 1942.

\bibitem{shaw2009wound}
Tanya~J Shaw and Paul Martin.
\newblock Wound repair at a glance.
\newblock {\em Journal of cell science}, 122(18):3209--3213, 2009.

\bibitem{abercrombie1970contact}
M~Abercrombie.
\newblock Contact inhibition in tissue culture.
\newblock {\em In vitro}, 6(2):128--142, 1970.

\bibitem{folkman1978role}
Judah Folkman and Anne Moscona.
\newblock Role of cell shape in growth control.
\newblock {\em Nature}, 273(5661):345, 1978.

\bibitem{chen1997geometric}
Christopher~S Chen, Milan Mrksich, Sui Huang, George~M Whitesides, and Donald~E
  Ingber.
\newblock Geometric control of cell life and death.
\newblock {\em Science}, 276(5317):1425--1428, 1997.

\bibitem{shraiman2005mechanical}
Boris~I Shraiman.
\newblock Mechanical feedback as a possible regulator of tissue growth.
\newblock {\em Proceedings of the National Academy of Sciences of the United
  States of America}, 102(9):3318--3323, 2005.

\bibitem{streichan2014spatial}
Sebastian~J Streichan, Christian~R Hoerner, Tatjana Schneidt, Daniela Holzer,
  and Lars Hufnagel.
\newblock Spatial constraints control cell proliferation in tissues.
\newblock {\em Proceedings of the National Academy of Sciences},
  111(15):5586--5591, 2014.

\bibitem{jacobeen2018cellular}
Shane Jacobeen, Jennifer~T Pentz, Elyes~C Graba, Colin~G Brandys, William~C
  Ratcliff, and Peter~J Yunker.
\newblock Cellular packing, mechanical stress and the evolution of
  multicellularity.
\newblock {\em Nature Physics}, 14(3):286, 2018.

\bibitem{puliafito2012collective}
Alberto Puliafito, Lars Hufnagel, Pierre Neveu, Sebastian Streichan, Alex
  Sigal, D~Kuchnir Fygenson, and Boris~I Shraiman.
\newblock Collective and single cell behavior in epithelial contact inhibition.
\newblock {\em Proceedings of the National Academy of Sciences},
  109(3):739--744, 2012.

\bibitem{mcclatchey2012contact}
Andrea~I McClatchey and Alpha~S Yap.
\newblock Contact inhibition (of proliferation) redux.
\newblock {\em Current opinion in cell biology}, 24(5):685--694, 2012.

\bibitem{kourtidis2017central}
Antonis Kourtidis, Ruifeng Lu, Lindy~J Pence, and Panos~Z Anastasiadis.
\newblock A central role for cadherin signaling in cancer.
\newblock {\em Experimental cell research}, 358(1):78--85, 2017.

\bibitem{irvine2017mechanical}
Kenneth~D Irvine and Boris~I Shraiman.
\newblock Mechanical control of growth: ideas, facts and challenges.
\newblock {\em Development}, 144(23):4238--4248, 2017.

\bibitem{vleminckx1991genetic}
Kris Vleminckx, Luc Vakaet, Marcus Mareel, Walter Fiers, and Frans Van~Roy.
\newblock {Genetic manipulation of E-cadherin expression by epithelial tumor
  cells reveals an invasion suppressor role}.
\newblock {\em Cell}, 66(1):107--119, 1991.

\bibitem{perl1998causal}
Anne-Karina Perl, Petra Wilgenbus, Ulf Dahl, Henrik Semb, and Gerhard
  Christofori.
\newblock {A causal role for E-cadherin in the transition from adenoma to
  carcinoma}.
\newblock {\em Nature}, 392(6672):190, 1998.

\bibitem{frixen1991cadherin}
Uwe~H Frixen, Jfirgen Behrens, Martin Sachs, Gertrud Eberle, Beate Voss,
  Angelika Warda, Dorothea L{\"o}chner, and Walter Birchmeier.
\newblock E-cadherin-mediated cell-cell adhesion prevents invasiveness of human
  carcinoma cells.
\newblock {\em The Journal of cell biology}, 113(1):173--185, 1991.

\bibitem{wong2003adhesion}
Alice~ST Wong and Barry~M Gumbiner.
\newblock {Adhesion-independent mechanism for suppression of tumor cell
  invasion by E-cadherin}.
\newblock {\em The Journal of cell biology}, 161(6):1191--1203, 2003.

\bibitem{gumbiner1996cell}
Barry~M Gumbiner.
\newblock Cell adhesion: the molecular basis of tissue architecture and
  morphogenesis.
\newblock {\em Cell}, 84(3):345--357, 1996.

\bibitem{cano2000transcription}
Amparo Cano, Mirna~A P{\'e}rez-Moreno, Isabel Rodrigo, Annamaria Locascio,
  Mar{\'i}a~J Blanco, Marta~G del Barrio, Francisco Portillo, and M~Angela
  Nieto.
\newblock The transcription factor snail controls epithelial-mesenchymal
  transitions by repressing {E}-cadherin expression.
\newblock {\em Nature cell biology}, 2(2):76, 2000.

\bibitem{yang2008epithelial}
Jing Yang and Robert~A Weinberg.
\newblock Epithelial-mesenchymal transition: at the crossroads of development
  and tumor metastasis.
\newblock {\em Developmental cell}, 14(6):818--829, 2008.

\bibitem{weinberg2013biology}
Robert Weinberg.
\newblock {\em The biology of cancer}.
\newblock Garland science, New York, 2013.

\bibitem{lou2008epithelial}
Yuanmei Lou, Olena Preobrazhenska, Margaret Sutcliffe, Lorena Barclay, Paul~C
  McDonald, Calvin Roskelley, Christopher~M Overall, Shoukat Dedhar, et~al.
\newblock Epithelial-mesenchymal transition ({EMT}) is not sufficient for
  spontaneous murine breast cancer metastasis.
\newblock {\em Developmental Dynamics}, 237(10):2755--2768, 2008.

\bibitem{fischer2015emt}
Kari~R Fischer, Anna Durrans, Sharrell Lee, Jianting Sheng, Fuhai Li,
  Stephen~TC Wong, Hyejin Choi, Tina El~Rayes, Seongho Ryu, Juliane Troeger,
  et~al.
\newblock Epithelial-to-mesenchymal transition is not required for lung
  metastasis but contributes to chemoresistance.
\newblock {\em Nature}, 527(7579):472, 2015.

\bibitem{zheng2015emt}
Xiaofeng Zheng, Julienne~L Carstens, Jiha Kim, Matthew Scheible, Judith Kaye,
  Hikaru Sugimoto, Chia-Chin Wu, Valerie~S LeBleu, and Raghu Kalluri.
\newblock Epithelial-to-mesenchymal transition is dispensable for metastasis
  but induces chemoresistance in pancreatic cancer.
\newblock {\em Nature}, 527(7579):525, 2015.

\bibitem{cheung2013collective}
Kevin~J Cheung, Edward Gabrielson, Zena Werb, and Andrew~J Ewald.
\newblock Collective invasion in breast cancer requires a conserved basal
  epithelial program.
\newblock {\em Cell}, 155(7):1639--1651, 2013.

\bibitem{shamir2014twist1}
Eliah~R Shamir, Elisa Pappalardo, Danielle~M Jorgens, Kester Coutinho, Wen-Ting
  Tsai, Khaled Aziz, Manfred Auer, Phuoc~T Tran, Joel~S Bader, and Andrew~J
  Ewald.
\newblock Twist1-induced dissemination preserves epithelial identity and
  requires {E}-cadherin.
\newblock {\em J Cell Biol}, 204(5):839--856, 2014.

\bibitem{silvera2009essential}
Deborah Silvera, Rezina Arju, Farbod Darvishian, Paul~H Levine, Ladan
  Zolfaghari, Judith Goldberg, Tsivia Hochman, Silvia~C Formenti, and Robert~J
  Schneider.
\newblock {Essential role for eIF4GI overexpression in the pathogenesis of
  inflammatory breast cancer}.
\newblock {\em Nature cell biology}, 11(7):903, 2009.

\bibitem{lewis2010misregulated}
Laura~J Lewis-Tuffin, Fausto Rodriguez, Caterina Giannini, Bernd Scheithauer,
  Brian~M Necela, Jann~N Sarkaria, and Panos~Z Anastasiadis.
\newblock {Misregulated E-cadherin expression associated with an aggressive
  brain tumor phenotype}.
\newblock {\em PloS one}, 5(10):e13665, 2010.

\bibitem{rodriguez2012cadherin}
Fausto~J Rodriguez, Laura~J Lewis-Tuffin, and Panos~Z Anastasiadis.
\newblock E-cadherin's dark side: possible role in tumor progression.
\newblock {\em Biochimica et Biophysica Acta (BBA)-Reviews on Cancer},
  1826(1):23--31, 2012.

\bibitem{padmanaban2019cadherin}
Veena Padmanaban, Ilona Krol, Yasir Suhail, Barbara~M Szczerba, Nicola Aceto,
  Joel~S Bader, and Andrew~J Ewald.
\newblock E-cadherin is required for metastasis in multiple models of breast
  cancer.
\newblock {\em Nature}, 573(7774):439--444, 2019.

\bibitem{liu2006cadherin}
Wendy~F Liu, Celeste~M Nelson, Dana~M Pirone, and Christopher~S Chen.
\newblock E-cadherin engagement stimulates proliferation via {R}ac1.
\newblock {\em The Journal of cell biology}, 173(3):431--441, 2006.

\bibitem{gray2008engineering}
Darren~S Gray, Wendy~F Liu, Colette~J Shen, Kiran Bhadriraju, Celeste~M Nelson,
  and Christopher~S Chen.
\newblock Engineering amount of cell--cell contact demonstrates biphasic
  proliferative regulation through {R}ho{A} and the actin cytoskeleton.
\newblock {\em Experimental cell research}, 314(15):2846--2854, 2008.

\bibitem{yap2015adherens}
Alpha~S Yap, Guillermo~A Gomez, and Robert~G Parton.
\newblock Adherens junctions revisualized: Organizing cadherins as
  nanoassemblies.
\newblock {\em Developmental cell}, 35(1):12--20, 2015.

\bibitem{amack2012knowing}
Jeffrey~D Amack and M~Lisa Manning.
\newblock Knowing the boundaries: extending the differential adhesion
  hypothesis in embryonic cell sorting.
\newblock {\em Science}, 338(6104):212--215, 2012.

\bibitem{david2014tissue}
Robert David, Olivia Luu, Erich~W Damm, Jason~WH Wen, Martina Nagel, and Rudolf
  Winklbauer.
\newblock Tissue cohesion and the mechanics of cell rearrangement.
\newblock {\em Development}, 141(19):3672--3682, 2014.

\bibitem{baronsky2016reduction}
Thilo Baronsky, Aliaksandr Dzementsei, Marieelen Oelkers, Juliane Melchert,
  Tomas Pieler, and Andreas Janshoff.
\newblock Reduction in {E}-cadherin expression fosters migration of {X}enopus
  laevis primordial germ cells.
\newblock {\em Integrative Biology}, 8(3):349--358, 2016.

\bibitem{valencia2015collective}
Angela M~Jimenez Valencia, Pei-Hsun Wu, Osman~N Yogurtcu, Pranay Rao, Josh
  DiGiacomo, In{\^e}s Godet, Lijuan He, Meng-Horng Lee, Daniele Gilkes, Sean~X
  Sun, et~al.
\newblock Collective cancer cell invasion induced by coordinated contractile
  stresses.
\newblock {\em Oncotarget}, 6(41):43438, 2015.

\bibitem{van2008cell}
Frans Van~Roy and Geert Berx.
\newblock The cell-cell adhesion molecule {E}-cadherin.
\newblock {\em Cellular and molecular life sciences}, 65(23):3756--3788, 2008.

\bibitem{kamkin2005mechanosensitivity}
Andre~Glebovich Kamkin and Irina~S Kiseleva.
\newblock {\em Mechanosensitivity in cells and tissues}.
\newblock Springer, 2005.

\bibitem{dolega2017cell}
ME~Dolega, Morgan Delarue, Fran{\c{c}}ois Ingremeau, Jacques Prost, Antoine
  Delon, and Giovanni Cappello.
\newblock Cell-like pressure sensors reveal increase of mechanical stress
  towards the core of multicellular spheroids under compression.
\newblock {\em Nature communications}, 8:14056, 2017.

\bibitem{mongera2018fluid}
Alessandro Mongera, Payam Rowghanian, Hannah~J Gustafson, Elijah Shelton,
  David~A Kealhofer, Emmet~K Carn, Friedhelm Serwane, Adam~A Lucio, James
  Giammona, and Otger Camp{\`a}s.
\newblock A fluid-to-solid jamming transition underlies vertebrate body axis
  elongation.
\newblock {\em Nature}, 561(7723):401, 2018.

\bibitem{dimilla1991mathematical}
PA~DiMilla, Kenneth Barbee, and DA~Lauffenburger.
\newblock Mathematical model for the effects of adhesion and mechanics on cell
  migration speed.
\newblock {\em Biophysical journal}, 60(1):15--37, 1991.

\bibitem{dimilla1993maximal}
Paul~A DiMilla, Julie~A Stone, John~A Quinn, Steven~M Albelda, and Douglas~A
  Lauffenburger.
\newblock Maximal migration of human smooth muscle cells on fibronectin and
  type {IV} collagen occurs at an intermediate attachment strength.
\newblock {\em The Journal of cell biology}, 122(3):729--737, 1993.

\bibitem{ahmadzadeh2017modeling}
Hossein Ahmadzadeh, Marie~R Webster, Reeti Behera, Angela M~Jimenez Valencia,
  Denis Wirtz, Ashani~T Weeraratna, and Vivek~B Shenoy.
\newblock Modeling the two-way feedback between contractility and matrix
  realignment reveals a nonlinear mode of cancer cell invasion.
\newblock {\em Proceedings of the National Academy of Sciences},
  114(9):E1617--E1626, 2017.

\bibitem{klank2017biphasic}
Rebecca~L Klank, Stacy A~Decker Grunke, Benjamin~L Bangasser, Colleen~L
  Forster, Matthew~A Price, David~J Odde, Karen~S SantaCruz, Steven~S
  Rosenfeld, Peter Canoll, Eva~A Turley, et~al.
\newblock Biphasic dependence of glioma survival and cell migration on {CD}44
  expression level.
\newblock {\em Cell reports}, 18(1):23--31, 2017.

\bibitem{li2003trends}
Christopher~I Li, Benjamin~O Anderson, Janet~R Daling, and Roger~E Moe.
\newblock Trends in incidence rates of invasive lobular and ductal breast
  carcinoma.
\newblock {\em Jama}, 289(11):1421--1424, 2003.

\bibitem{richardson2010mechanisms}
Brian~E Richardson and Ruth Lehmann.
\newblock Mechanisms guiding primordial germ cell migration: strategies from
  different organisms.
\newblock {\em Nature reviews Molecular cell biology}, 11(1):37, 2010.

\bibitem{mendonsa2018ecadherin}
Alisha~M Mendonsa, Tae-Young Na, and Barry~M Gumbiner.
\newblock E-cadherin in contact inhibition and cancer.
\newblock {\em Oncogene}, 37:4769--4780, 2018.

\bibitem{brouxhon2013monoclonal}
Sabine~M Brouxhon, Stephanos Kyrkanides, Xiaofei Teng, Veena Raja, M~Kerry
  O'Banion, Robert Clarke, Stephen Byers, Andrew Silberfeld, Carmen Tornos, and
  Li~Ma.
\newblock {Monoclonal antibody against the ectodomain of E-cadherin (DECMA-1)
  suppresses breast carcinogenesis: involvement of the HER/PI3K/Akt/mTOR and
  IAP pathways}.
\newblock {\em Clinical Cancer Research}, 19(12):3234--3246, 2013.

\bibitem{carneiro2013therapeutic}
Patr{\'\i}cia Carneiro, Joana Figueiredo, Renata Bordeira-Carri{\c{c}}o,
  Maria~Sofia Fernandes, Joana Carvalho, Carla Oliveira, and Raquel Seruca.
\newblock Therapeutic targets associated to {E}-cadherin dysfunction in gastric
  cancer.
\newblock {\em Expert opinion on therapeutic targets}, 17(10):1187--1201, 2013.

\bibitem{levayer2016tissue}
Romain Levayer, Carole Dupont, and Eduardo Moreno.
\newblock Tissue crowding induces caspase-dependent competition for space.
\newblock {\em Current Biology}, 26(5):670--677, 2016.

\bibitem{schaller2005multicellular}
Gernot Schaller and Michael Meyer-Hermann.
\newblock {Multicellular tumor spheroid in an off-lattice Voronoi-Delaunay cell
  model}.
\newblock {\em Physical Review E}, 71(5):051910, 2005.

\bibitem{drasdo2005single}
Dirk Drasdo and Stefan H{\"o}hme.
\newblock A single-cell-based model of tumor growth in vitro: monolayers and
  spheroids.
\newblock {\em Physical biology}, 2(3):133, 2005.

\bibitem{galle2005modeling}
J{\"o}rg Galle, Markus Loeffler, and Dirk Drasdo.
\newblock Modeling the effect of deregulated proliferation and apoptosis on the
  growth dynamics of epithelial cell populations in vitro.
\newblock {\em Biophysical journal}, 88(1):62--75, 2005.

\bibitem{malmi2018cell}
Abdul~N Malmi-Kakkada, Xin Li, Himadri~S Samanta, Sumit Sinha, and Dave
  Thirumalai.
\newblock Cell growth rate dictates the onset of glass to fluidlike transition
  and long time superdiffusion in an evolving cell colony.
\newblock {\em Physical Review X}, 8(2):021025, 2018.

\bibitem{pathmanathan2009computational}
P~Pathmanathan, Jonathan Cooper, Alexander Fletcher, Gary Mirams, P~Murray,
  J~Osborne, Joe Pitt-Francis, A~Walter, and SJ~Chapman.
\newblock A computational study of discrete mechanical tissue models.
\newblock {\em Physical biology}, 6(3):036001, 2009.

\bibitem{brodland2002differential}
G~Wayne Brodland.
\newblock {The differential interfacial tension hypothesis (DITH): a
  comprehensive theory for the self-rearrangement of embryonic cells and
  tissues}.
\newblock {\em Journal of biomechanical engineering}, 124(2):188--197, 2002.

\bibitem{hayashi2004surface}
Takashi Hayashi and Richard~W Carthew.
\newblock Surface mechanics mediate pattern formation in the developing retina.
\newblock {\em Nature}, 431(7009):647--652, 2004.

\bibitem{conger}
Alan~D Conger and Marvin~C Ziskin.
\newblock Growth of mammalian multicellular tumor spheroids.
\newblock {\em Cancer Res}, 43(2):556--560, 1983.

\bibitem{Mandonnet2003}
Emmanuel Mandonnet, Jean-Yves Delattre, Marie-Laure Tanguy, Kristin~R Swanson,
  Antoine~F Carpentier, Hugues Duffau, Philippe Cornu, R{\'e}my Van~Effenterre,
  Ellsworth~C Alvord, and Laurent Capelle.
\newblock {Continuous growth of mean tumor diameter in a subset of grade II
  gliomas}.
\newblock {\em Ann Neurol}, 53(4):524--528, 2003.

\bibitem{Simeoni2004}
Monica Simeoni, Paolo Magni, Cristiano Cammia, Giuseppe De~Nicolao, Valter
  Croci, Enrico Pesenti, Massimiliano Germani, Italo Poggesi, and Maurizio
  Rocchetti.
\newblock Predictive pharmacokinetic-pharmacodynamic modeling of tumor growth
  kinetics in xenograft models after administration of anticancer agents.
\newblock {\em Cancer Res}, 64(3):1094--1101, 2004.

\bibitem{Hart1998}
D~Hart, E~Shochat, and Z~Agur.
\newblock The growth law of primary breast cancer as inferred from mammography
  screening trials data.
\newblock {\em Br J Cancer}, 78(3):382--387, 1998.

\bibitem{grimes}
David~Robert Grimes, Pavitra Kannan, Alan McIntyre, Anthony Kavanagh, Abul
  Siddiky, Simon Wigfield, Adrian Harris, and Mike Partridge.
\newblock The role of oxygen in avascular tumor growth.
\newblock {\em PloS one}, 11(4):e0153692, 2016.

\bibitem{helmlinger1997solid}
Gabriel Helmlinger, Paolo~A Netti, Hera~C Lichtenbeld, Robert~J Melder, and
  Rakesh~K Jain.
\newblock Solid stress inhibits the growth of multicellular tumor spheroids.
\newblock {\em Nature biotechnology}, 15(8):778--783, 1997.

\bibitem{eagle1967growth}
Harry Eagle and Elliot~M Levine.
\newblock Growth regulatory effects of cellular interaction.
\newblock {\em Nature}, 213(5081):1102--1106, 1967.

\bibitem{dallon}
John~C Dallon and Hans~G Othmer.
\newblock {How cellular movement determines the collective force generated by
  the Dictyostelium discoideum slug}.
\newblock {\em Journal of theoretical biology}, 231(2):203--222, 2004.

\bibitem{samanta2019origin}
Himadri~S Samanta and D~Thirumalai.
\newblock Origin of superdiffusive behavior in a class of nonequilibrium
  systems.
\newblock {\em Physical Review E}, 99(3):032401, 2019.

\bibitem{Dammer95Science}
U~Dammer, O~Popescu, P~Wagner, D~Anselmetti, HJ~Guntherodt, and GN~Misevic.
\newblock Binding strength between cell-adhesion proteoglycans measured by
  atomic-force microscopy.
\newblock {\em {Science}}, {267}({5201}):{1173--1175}, {1995}.

\bibitem{gibson2006emergence}
Matthew~C Gibson, Ankit~B Patel, Radhika Nagpal, and Norbert Perrimon.
\newblock The emergence of geometric order in proliferating metazoan epithelia.
\newblock {\em Nature}, 442(7106):1038, 2006.

\bibitem{fischer2017three}
Sabine~C Fischer, Elena Corujo-Simon, Joaquin Lilao-Garzon, Ernst~HK Stelzer,
  and Silvia Munoz-Descalzo.
\newblock Three-dimensional cell neighbourhood impacts differentiation in the
  inner mass cells of the mouse blastocyst.
\newblock {\em bioRxiv}, page 159301, 2017.

\bibitem{byers1995role}
Stephen~W Byers, Connie~L Sommers, Becky Hoxter, Arthur~M Mercurio, and Aydin
  Tozeren.
\newblock {Role of E-cadherin in the response of tumor cell aggregates to
  lymphatic, venous and arterial flow: measurement of cell-cell adhesion
  strength}.
\newblock {\em Journal of cell science}, 108(5):2053--2064, 1995.

\end{thebibliography}


\begin{thebibliography}{10}

\bibitem{irving1950statistical}
JH~Irving and John~G Kirkwood.
\newblock {The statistical mechanical theory of transport processes. IV. The
  equations of hydrodynamics}.
\newblock {\em The Journal of chemical physics}, 18(6):817--829, 1950.

\bibitem{yang2012generalized}
Jerry~Zhijian Yang, Xiaojie Wu, and Xiantao Li.
\newblock {A generalized Irving--Kirkwood formula for the calculation of stress
  in molecular dynamics models}.
\newblock {\em The Journal of chemical physics}, 137(13):134104, 2012.

\bibitem{ramis2008modeling}
Ignacio Ramis-Conde, Dirk Drasdo, Alexander~RA Anderson, and Mark~AJ Chaplain.
\newblock {Modeling the influence of the E-cadherin-$\beta$-catenin pathway in
  cancer cell invasion: a multiscale approach}.
\newblock {\em Biophysical journal}, 95(1):155--165, 2008.

\bibitem{berx2001cadherin}
Geert Berx and Frans Van~Roy.
\newblock {The E-cadherin/catenin complex: an important gatekeeper in breast
  cancer tumorigenesis and malignant progression}.
\newblock {\em Breast Cancer Research}, 3(5):289, 2001.

\bibitem{rodriguez2012cadherin}
Fausto~J Rodriguez, Laura~J Lewis-Tuffin, and Panos~Z Anastasiadis.
\newblock E-cadherin's dark side: possible role in tumor progression.
\newblock {\em Biochimica et Biophysica Acta (BBA)-Reviews on Cancer},
  1826(1):23--31, 2012.

\bibitem{tan1999biological}
DS~Tan, HW~Potts, AC~Leong, CE~Gillett, D~Skilton, W~Hetal Harris, RD~Liebmann,
  and AM~Hanby.
\newblock {The biological and prognostic significance of cell polarity and
  E-cadherin in grade I infiltrating ductal carcinoma of the breast}.
\newblock {\em The Journal of pathology}, 189(1):20--27, 1999.

\bibitem{kleer2001persistent}
Celina~G Kleer, Kenneth~L van Golen, Thomas Braun, and Sofia~D Merajver.
\newblock {Persistent E-cadherin expression in inflammatory breast cancer}.
\newblock {\em Modern Pathology}, 14(5):458, 2001.

\bibitem{lewis2010misregulated}
Laura~J Lewis-Tuffin, Fausto Rodriguez, Caterina Giannini, Bernd Scheithauer,
  Brian~M Necela, Jann~N Sarkaria, and Panos~Z Anastasiadis.
\newblock {Misregulated E-cadherin expression associated with an aggressive
  brain tumor phenotype}.
\newblock {\em PloS one}, 5(10):e13665, 2010.

\bibitem{utsuki2002relationship}
Satoshi Utsuki, Yuichi Sato, Hidehiro Oka, Benio Tsuchiya, Sachio Suzuki, and
  Kiyotaka Fujii.
\newblock {Relationship between the expression of E-, N-cadherins and
  beta-catenin and tumor grade in astrocytomas}.
\newblock {\em Journal of neuro-oncology}, 57(3):187--192, 2002.

\bibitem{perego2002invasive}
Carla Perego, Cristina Vanoni, Silvia Massari, Andrea Raimondi, Sandra Pola,
  Maria~Grazia Cattaneo, Maura Francolini, Lucia~Maria Vicentini, and Grazia
  Pietrini.
\newblock Invasive behaviour of glioblastoma cell lines is associated with
  altered organisation of the cadherin-catenin adhesion system.
\newblock {\em Journal of Cell Science}, 115(16):3331--3340, 2002.

\bibitem{gofuku1999expression}
Junji Gofuku, Hitoshi Shiozaki, Toshimasa Tsujinaka, Masatoshi Inoue, Shigeyuki
  Tamura, Yuichiro Doki, Shigeo Matsui, Shoichiro Tsukita, Nobuteru Kikkawa,
  and Morito Monden.
\newblock {Expression of E-cadherin and $\alpha$-catenin in patients with
  colorectal carcinoma: correlation with cancer invasion and metastasis}.
\newblock {\em American journal of clinical pathology}, 111(1):29--37, 1999.

\bibitem{dorudi1993cadherin}
S~Dorudi, JP~Sheffield, R~Poulsom, JM~Northover, and IR~Hart.
\newblock {E-cadherin expression in colorectal cancer. An immunocytochemical
  and in situ hybridization study.}
\newblock {\em The American journal of pathology}, 142(4):981, 1993.

\bibitem{sundfeldt1997cadherin}
Karin Sundfeldt, Yael Piontkewitz, Karin Ivarsson, Ola Nilsson, P{\"a}r
  Hellberg, Mats Br{\"a}nnstr{\"o}, Per-Olof Janson, Sven Enerb{\"a}ck, Lars
  Hedin, et~al.
\newblock E-cadherin expression in human epithelial ovarian cancer and normal
  ovary.
\newblock {\em International journal of cancer}, 74(3):275--280, 1997.

\bibitem{reddy2005formation}
Pradeep Reddy, Lian Liu, Chong Ren, Peter Lindgren, Karin Boman, Yan Shen, Eva
  Lundin, Ulrika Ottander, Miia Rytinki, and Kui Liu.
\newblock {Formation of E-cadherin-mediated cell-cell adhesion activates AKT
  and mitogen activated protein kinase via phosphatidylinositol 3 kinase and
  ligand-independent activation of epidermal growth factor receptor in ovarian
  cancer cells}.
\newblock {\em Molecular Endocrinology}, 19(10):2564--2578, 2005.

\bibitem{kim2016loss}
Sun~A Kim, Kentaro Inamura, Mai Yamauchi, Reiko Nishihara, Kosuke Mima,
  Yasutaka Sukawa, Tingting Li, Mika Yasunari, Teppei Morikawa, Kathryn~C
  Fitzgerald, et~al.
\newblock {Loss of CDH1 (E-cadherin) expression is associated with infiltrative
  tumour growth and lymph node metastasis}.
\newblock {\em British journal of cancer}, 114(2):199, 2016.

\bibitem{malmi2018cell}
Abdul~N Malmi-Kakkada, Xin Li, Himadri~S Samanta, Sumit Sinha, and Dave
  Thirumalai.
\newblock Cell growth rate dictates the onset of glass to fluidlike transition
  and long time superdiffusion in an evolving cell colony.
\newblock {\em Physical Review X}, 8(2):021025, 2018.

\bibitem{schaller2005multicellular}
Gernot Schaller and Michael Meyer-Hermann.
\newblock {Multicellular tumor spheroid in an off-lattice Voronoi-Delaunay cell
  model}.
\newblock {\em Physical Review E}, 71(5):051910, 2005.

\bibitem{galle2005modeling}
J{\"o}rg Galle, Markus Loeffler, and Dirk Drasdo.
\newblock Modeling the effect of deregulated proliferation and apoptosis on the
  growth dynamics of epithelial cell populations in vitro.
\newblock {\em Biophysical journal}, 88(1):62--75, 2005.

\bibitem{freyer}
James~P Freyer and Robert~M Sutherland.
\newblock {Regulation of growth saturation and development of necrosis in
  EMT6/Ro multicellular spheroids by the glucose and oxygen supply}.
\newblock {\em Cancer research}, 46(7):3504--3512, 1986.

\bibitem{casciari}
Joseph~J Casciari, Stratis~V Sotirchos, and Robert~M Sutherland.
\newblock {Variations in tumor cell growth rates and metabolism with oxygen
  concentration, glucose concentration, and extracellular pH}.
\newblock {\em Journal of cellular physiology}, 151(2):386--394, 1992.

\bibitem{landry}
Jacques Landry, James~P Freyer, and Robert~M Sutherland.
\newblock Shedding of mitotic cells from the surface of multicell spheroids
  during growth.
\newblock {\em Journal of cellular physiology}, 106(1):23--32, 1981.

\bibitem{montel2011stress}
Fabien Montel, Morgan Delarue, Jens Elgeti, Laurent Malaquin, Markus Basan,
  Thomas Risler, Bernard Cabane, Danijela Vignjevic, Jacques Prost, Giovanni
  Cappello, et~al.
\newblock Stress clamp experiments on multicellular tumor spheroids.
\newblock {\em Physical review letters}, 107(18):188102, 2011.

\end{thebibliography}
\bibliographystyle{unsrt}

\end{document}

\title{Supplementary Information: Dual Role of Cell-Cell Adhesion In Tumor Suppression and Proliferation Due to Collective Mechanosensing}

\author{Abdul N Malmi-Kakkada $^{a}$, Xin Li $^{a}$, Sumit Sinha $^{b}$, D. Thirumalai $^{a}$} 
%author[a]{Author Three}

\affiliation{$^{a}$Department of Chemistry, University of Texas at Austin, Austin, TX 78712, \\  $^{b}$Department of Physics, University of Texas at Austin, Austin, TX 78712}
%\footnote{Corresponding author: dave.thirumalai@gmail.com}}

\vspace{2cm}
\date{\today}

\maketitle

\section*{I. Nonlinear proliferation behavior is robust to alternative values of the critical pressure}

The pressure experienced by the cell, $p_i = \sum_{j \in NN(i)} \frac{|F_{ij}|}{A_{ij}}$, considers 
the absolute value of the force $|{F_{ij}}|$ exerted on a cell $i$. The use of the absolute value 
ensures that both repulsive (positive) and adhesive (negative) contributions to the pressure are treated on an equal footing, 
given any fixed positive value of the critical pressure. 

In order to ascertain if the non-monotonic dependence of proliferation on $f^{ad}$ depends on 
exact values and definition of the critical pressure, we varied $p_{c}$. We also considered alternative definitions of the 
critical pressure because the precise calculation of pressure in systems that are far from equilibrium is not entirely clear. 
\par
\begin{figure}
%\begin{turn}{-90}
\centering
\includegraphics[width=0.7\textwidth] {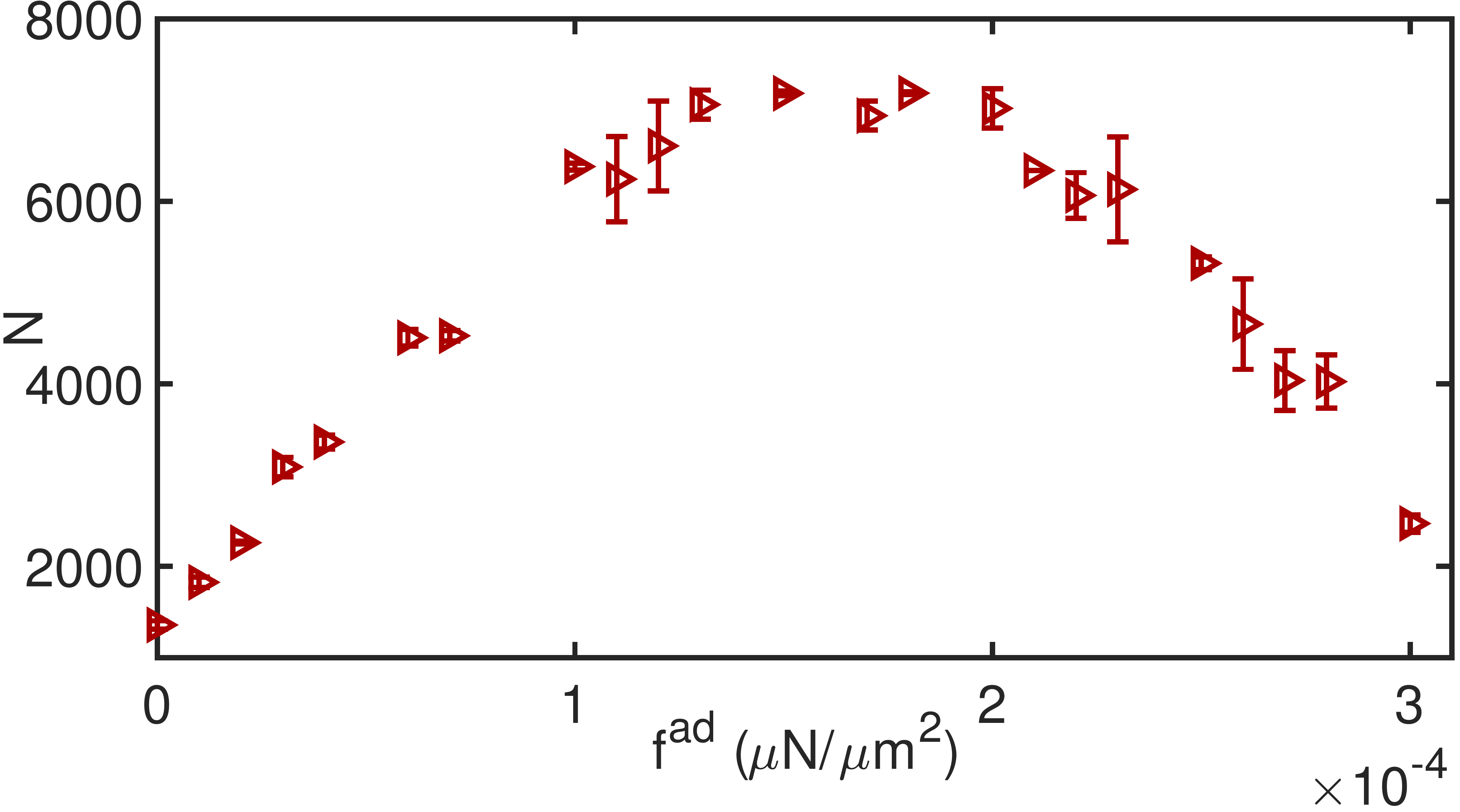} %oa
%\end{center}
%\end{turn}
\caption{Number of cells after 7.5 days of growth as a function of 
cell-cell adhesion strength at $p_c = 5\times 10^{-5}$ MPa, which is a factor of two less than the value (see Table S1) used in the Main Text.} %inset pressure figure showing h_0 would be appropriate here.
\label{figsi5}
\end{figure}

{\bf Role of p$_{c}$:} For $p_{c} = 5\times 10^{-5} \mathrm{MPa}$, the size of the spheroid,  
N(t=7.5 days), is shown in Fig.~\ref{figsi5}. 
The biphasic behavior also persists at the lower critical pressure of $p_c = 5\times 10^{-5}$ MPa, 
indicating that the non-monotonic behavior of the proliferative capacity as a function of $f^{ad}$ 
does not depend on the exact value of $p_c$ in the range explored here. 
However, the proliferation extent measured by the total number of 
cells obtained after 7.5 days ($\sim 12\tau_{min})$ of growth, at 
all values of $f^{ad}$, is greatly reduced  
at lower $p_{c}$ (compare to Fig. 2a in the Main Text). For example, at $f^{ad}=1.5\times 10^{-4} \mathrm{\mu N/\mu m^2}$
(for $p_c = 5\times 10^{-5} \mathrm{MPa}$), there is a $\approx 40\%$ 
reduction in $N$ as compared to $p_c =1\times 10^{-4} \mathrm{MPa}$ (at fixed $t$). 
Lower critical pressure makes it easier for cells to 
enter the dormant state, leading to the inhibition of overall proliferation while preserving the variation of the size of the 
tumor as a function of $f^{ad}$. 

\par
{\bf Irving-Kirkwood Pressure:} Pressure experienced by cells 
can also be calculated based on the Irving-Kirkwood (IK) stress tensor. In systems out of equilibrium, it is beneficial 
to calculate local stress at certain points in space and time. %In such scenarios, local stress can be calculated based on the IK formalism. 
The IK stress tensor~\cite{irving1950statistical,yang2012generalized}, $\sigma^{\alpha' \beta'}$, is defined as, 
\begin{equation}
\sigma^{\alpha' \beta'}_{i} = \frac{1}{V}  \sum_{i \ne j \in NN(i)} F^{\alpha'}_{ij} |r^{\beta'}_{i}-r^{\beta'}_{j}| , 
\label{pressik}
\end{equation}
where $\alpha', \beta'=[x,y,z]$, $V=(4/3)\pi R_{i}^3+\sum_{j \in NN(i)}(4/3) \pi R_{j}^3$ is the local volume occupied by a cell 
and its nearest neighbors, and $F_{ij}$ is the magnitude of the force exerted on cell $i$ due to cell $j$. Here, the nearest neighbors 
of a cell $i$ ($NN(i)$) is defined as any cell $j$ with $h_{ij} > 0$. 
The IK pressure is the trace of the stress tensor $p^{IK}_{i}=\sigma^{\alpha' \alpha'}/3$. 

\begin{figure}
%\begin{turn}{-90}
\centering
\includegraphics[width=0.6\textwidth] {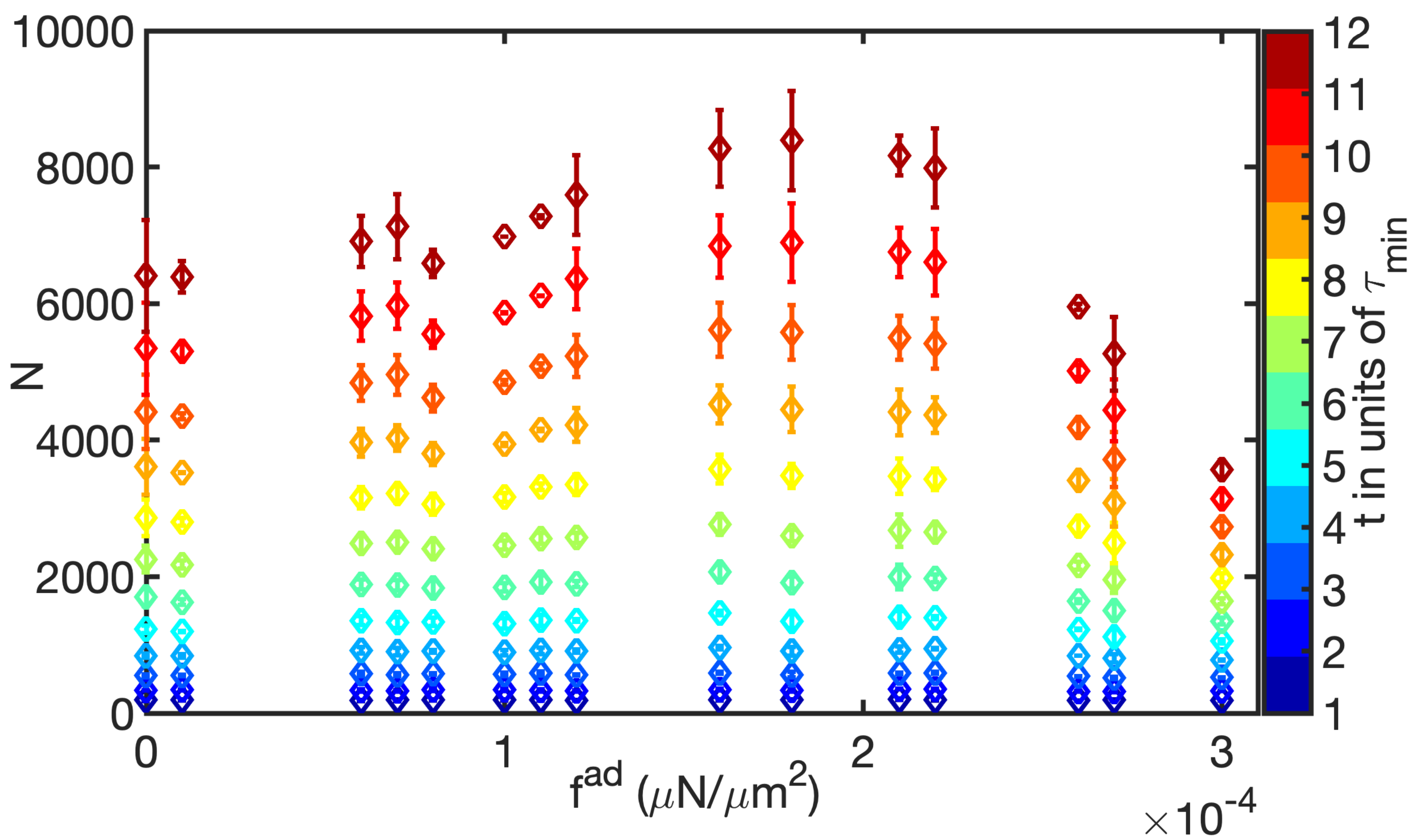} %oa
%\end{center}
%\end{turn}
\caption{Number of cells, at $t=\tau_{min}$ to $9\tau_{min}$ as a function of $f^{ad}$. We used 
$p_c=1.5\times 10^{-7} \mathrm{MPa}$ in the simulations with the Irving-Kirkwood definition of pressure 
Eq.~\ref{pressik}. These simulations also show that the observed non-monotonic dependence of proliferation on $f^{ad}$ is robust.} %inset pressure figure showing h_0 would be appropriate here.
\label{pressaltf}
\end{figure}

Non-monotonic proliferation behavior is observed even when the IK definition of pressure (see Fig.~\ref{pressaltf}) 
is used. The peak in the number of cells is at $f^{ad}\sim 1.75 \times 10^{-4}$, which is fairly close to the value found in 
Fig. 2 of the Main Text. We find that the magnitude of the pressure 
calculated using Eq.~(\ref{pressik}) is less than the values obtained by 
calculating $p_i$ as described in the Main Text. As a result, we used 
lower values of $p_{c}$ in order to explore the dependence of the proliferation on the $f^{ad}$. 
Because $p_{c}$ is small, the number of cells obtained after 
$5.6$ days of growth is less than what is found in Fig. 2a of the 
Main Text.

\section*{II. Use of different form of \texorpdfstring{$f^{ad}$}{Lg} preserves non-monotonic proliferation}
\label{nonlinear}

\begin{figure}
%\begin{turn}{-90}
\centering
\includegraphics[width=0.7\textwidth] {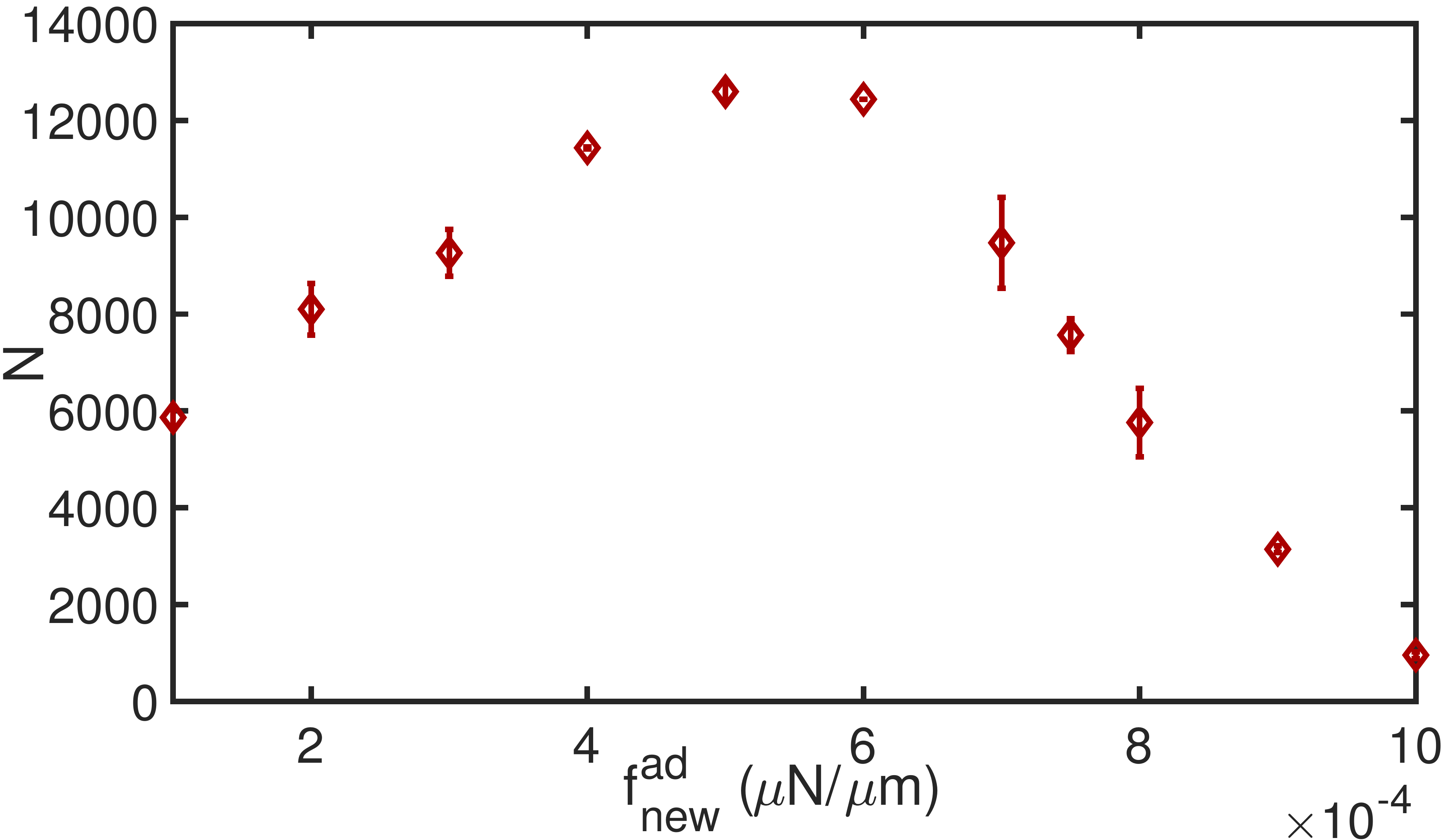} %oa
%\end{center}
%\end{turn}
\caption{Number of cells after 7.5 days of growth as a function of cell-cell adhesion strength 
using an alternate form of $F^{ad}=\rho_{m}W_{s}h_{ij}$. The results are qualitatively similar to the 
ones in Fig.~2 of the Main Text. } %inset pressure figure showing h_0 would be appropriate here.
\label{figsi4}
\end{figure}

Besides ensuring that the nonlinear proliferation behavior does not depend on the 
value of the critical pressure (see Fig.~\ref{figsi5}), 
we also tested whether our results are dependent on the form 
of the cell-cell interaction (see Eqs.~A3-~A4 Appendix A of the Main Text). %more general than the cell-cell repulsive and 
%adhesive interaction considered in the Main text . %Hence, we repeated our study 
%with a different form of adhesive interaction. 
We performed additional simulations using adhesive interaction of the form, 
\begin{equation}
\label{fneweq}
F^{ad} = \rho_{m}W_s h_{ij}, 
\end{equation}
where $\rho_m$ is the density of surface adhesion molecules, and 
$W_s$ is the adhesion energy of a single bond~\cite{ramis2008modeling}. 
%and $h_{ij}$ is as defined earlier. 
According to Eq.~\ref{fneweq}, 
with decreasing cell center-to-center distance 
or equivalently increasing cell-cell overlap $h_{ij}$, the number of adhesive contacts between cells increase.  
This leads to increased attractive interaction between the cells. The %modified Hertz-model based 
repulsive interaction is left unchanged. 
Defining a new cell-cell adhesion strength parameter, $f^{ad}_{new} =  \rho_{m}W_s$, 
the biphasic cell proliferation behavior is once again obtained (see Fig.~\ref{figsi4}).
Although the optimal cell-cell adhesion strength shifts to a different value, 
$f^{ad}_{opt,new}= 6\times 10^{-4}~\mathrm{\mu N/\mu m}$,  
compared to the interaction considered in the Main text, 
the overall trend is similar. Thus, the qualitative observation 
of non-monotonic dependence of proliferative capacity on $f^{ad}$ is unchanged, 
establishing the robustness of the results. 

\section*{III. Plausible connection between simulations and clinical data}

E-cadherin is considered to be primarily a tumor suppressor, based on the observation 
that it is down regulated 
during epithelial to mesenchymal transition (EMT)~\cite{berx2001cadherin,rodriguez2012cadherin}. %and its role in the formation of contacts between cells. 
The tumor suppressor role of E-cadherin (encoded by the CDH1 gene) 
has been elucidated in breast cancer where loss of heterozygosity in chromosome region 16q22.1 
(the gene region that codes for E-cadherin) is frequent~\cite{berx2001cadherin,rodriguez2012cadherin}. 
In recent years, however, an alternative role for E-cadherin as a tumor promoter seems to be emerging~\cite{tan1999biological,kleer2001persistent,rodriguez2012cadherin}. 

According to our findings (albeit using only simulations), the overall E-cadherin %(or any cell adhesion molecule - CAM - which mediates cell-cell adhesion) 
expression level determines its role as a tumor suppressor or promotor. 
For cells characterized by low/no E-cadherin expression, we hypothesize 
that increasing its expression leads to enhanced tumorigenicity, and by implication poor survival prognosis. 
On the other hand, if the native E-cadherin  
expression level is high, its up-regulation could lead to 
the suppression of tumor growth. %, and therefore better prospects for survival. 
In the context of cancer, heterogeneity in E-cadherin expression is observed. 
For example, E-cadherin 
expression is rare to non-existent in both brain tumors and normal brain tissues~\cite{lewis2010misregulated,utsuki2002relationship,perego2002invasive}. 
In colorectal tissues, however, epithelial cells express E-cadherin without exception~\cite{gofuku1999expression,dorudi1993cadherin}. 
%Using data mined from cBioportal (http://www.cbioportal.org)~\cite{cerami2012cbio,gao2013integrative}, 
%we analyze correlation between survival 
%and CDH1 mRNA expression for two types of cancers viz. glioblastoma multiforme 
%and colorectal adenocarcinoma selected for tissues with low and high 
%E-cadherin expression levels respectively. 
We should point out that it is difficult to quantitatively compare the key prediction made here 
(Fig.~2 in the Main Text) expressing the dual role of E-cadherin in tumor growth with available data. 
Nevertheless, it appears that there is evidence for the non-monotonic cell proliferation as a function 
of the strength of cell-cell attraction. 
We give anecdotal evidence for the dual role that E-cadherin plays in tumor evolution. 
\par
\begin{figure}
\centering
\includegraphics[width=\textwidth]{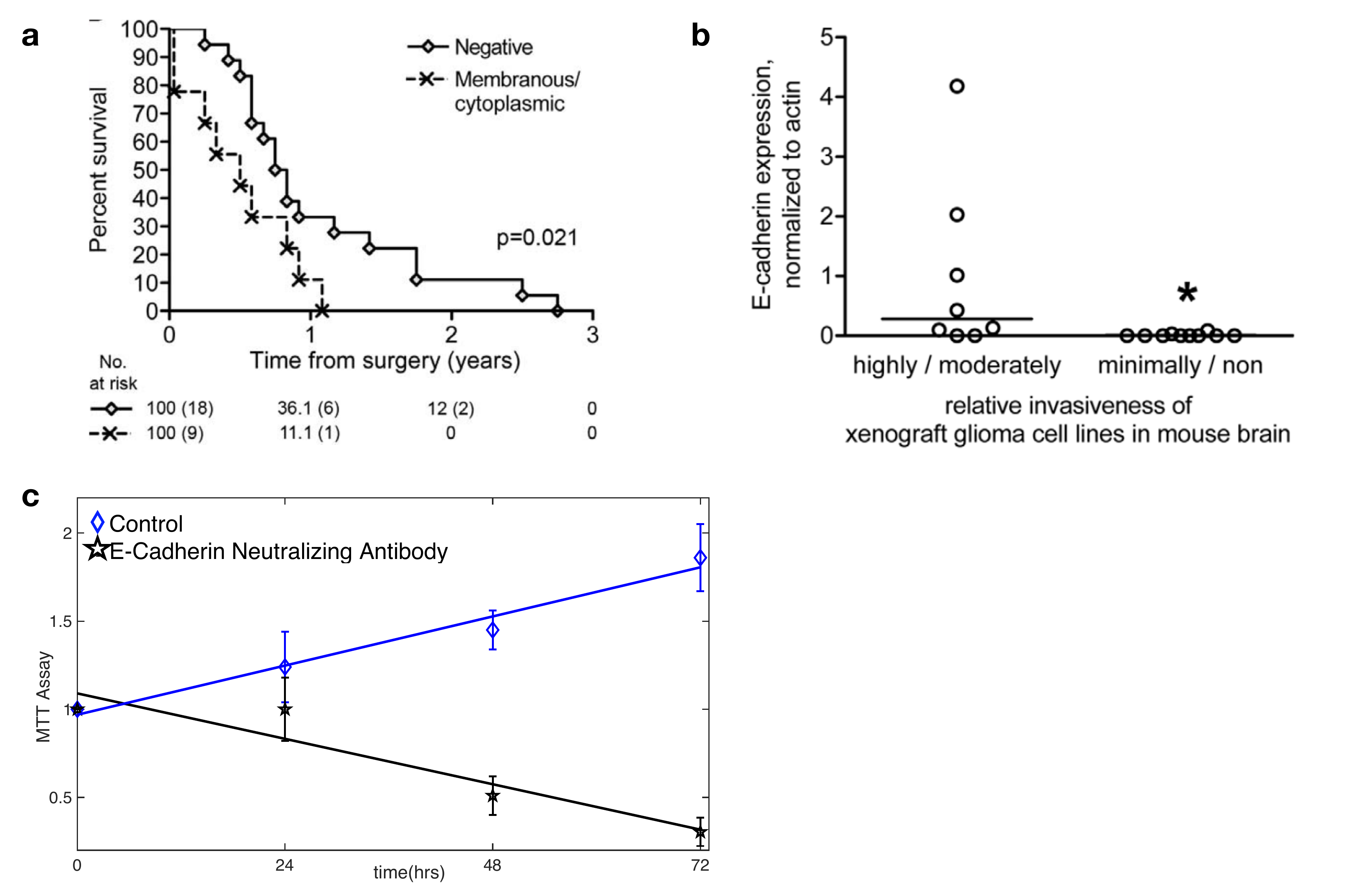}
\caption{Figs.~{\bf a)} $\&$ {\bf b)} are reproduced from Ref.~\cite{lewis2010misregulated}. 
{\bf a)} Overall survival from surgery for patients with tumors characterized by absence (Negative), presence (Membranous/cytoplasmic) 
E-cadherin expression. Patients whose tumors did not express E-cadherin had better overall 
survival compared to those that did express E-cadherin. Percent survival (No. of patients) are indicated at the bottom 
at different times from surgery. 
{\bf b)} Quantification of E-cadherin expression level compared to Actin versus relative invasiveness 
of xenograft GMB tumor in mouse brain.
{\bf c)} MTT Assay data for Ovarian tumor cells showing proliferation behavior for control cells expressing E-cadherin 
versus E-cadherin neutralized cells. The experimental results reproduced here give qualitative support to the simulation results.}
%{\bf d)}Sketch of proposed biphasic relationship between cell proliferation and cell-cell adhesion level. }
\label{figsi6}
\end{figure}

 \medskip
{\bf E-cadherin expression correlates with worse prognosis for glioblastoma and ovarian cancer:}
E-cadherin levels in tumor tissue samples from 27 individuals with a rare subtype of Glioblastoma Multiforme (GBM) with epithelial/
pseudoepithelial differentiation was analyzed by Lewis-Tuffin et. al.~\cite{lewis2010misregulated}. 
Nine out of the 27 cases exhibited E-cadherin expression. These patients demonstrated poorer overall survival 
compared to the 18 patients whose tumors did not express E-cadherin (see Fig.~\ref{figsi6}a, Negative stands for no E-cadherin expression compared 
to tumor cells exhibiting Membranous/cytoplasmic E-cadherin expression). 

After establishing orthotropic xenografts in mice, Lewis-Tuffin et.al~\cite{lewis2010misregulated} 
sectioned the brains to determine the relative invasiveness of the tumors 
depending on E-cadherin expression. Five out of the eight high/moderately invasive tumors 
expressed enhanced levels of E-cadherin while none of the minimally invasive tumors showed E-cadherin 
expression (see Fig.~\ref{figsi6}b). The data indicate that higher E-cadherin expression could be one of the 
contributors to GBM tumor aggressiveness. 

Similar to GBM, E-cadherin expression in ovarian cancers exhibits a distinct pattern. 
Healthy ovarian surface epithelial (OSE) cells do not express E-cadherin. However, it is  
consistently expressed in benign, borderline and malignant ovarian tumors at all stages, including in 
metastases from such ovarian tumors~\cite{sundfeldt1997cadherin,reddy2005formation}.
To ascertain the physiological function of E-cadherin in ovarian tumor cells, 3-(4,5-dimethylthiazol-2-yl)-2,5-diphenyltetrazolium bromide (MTT) cell viability and proliferation assays 
were performed. MTT assay involved seeding of $2\times 10^{5}$ of ovarian cancer cell line (OVCAR-3) in 24-well plates 
with or without E-cadherin neutralizing antibody. Number of cells at time $0$ was defined as $1.0$, and fold reductions/increments 
at different time points is indicated as mean $\pm$ standard deviation~\cite{reddy2005formation}.  
Control group of cells show consistent proliferation while 
neutralization of E-cadherin function in cancer cells led to marked suppression of cell proliferation (Fig.~\ref{figsi6}c). 
Therefore, one could surmise that E-cadherin plays the role of a tumor promoter in certain forms of Ovarian cancers. 
%Using 591 Glioblastoma Multiforme and 629 Colorectal Adenocarcinoma patient samples in 
%The Cancer Genome Atlas (TCGA) provisional database accessed through cBioportal, we studied the trend 
%in patient survival vs. CDH1 mRNA expression. Overall survival (months) is the length of time from either 
%the date of cancer diagnosis or start of cancer treatment that a patient is alive. For mRNA expression data, 
%the relative expression of a gene compared to gene expression in a reference population is computed. 
%All samples that are diploid for the gene or normal samples or all profiled samples is the reference. 
%Z-score indicates gene expression as number of standard deviations away from the mean reference expression. 
%This measure is useful in determining whether a gene is up or down-regulated relative to normal or other tumor 
%samples~\cite{cerami2012cbio,gao2013integrative}. 
As mentioned earlier, mapping our findings to the specific cancer types discussed here 
requires more precise data analysis, which would require additional experiments and simulations. The qualitative 
similarity between our findings and in certain cancer types is encouraging. 

%In GBM tumor tissues, characterized by low levels of E-cadherin expression, 
%increasing E-cadherin expression correlated with poor survival. %A negative correlation between survival and E-cadherin expression exists. 
%As to the role of E-cadherin as a tumor suppressor, 
{\bf Tumor suppression:} In colorectal cancers,  
infiltrative tumor growth and lymph node metastasis are correlated with loss of E-cadherin expression~\cite{kim2016loss}, 
implying that enhanced $f^{ad}$ leads to tumor suppression.    
Given the heterogeneity associated with cancer cell properties, it is possible that cells within a single cancer type 
may exhibit a wide variation in cellular adhesion molecule expression levels. In such a scenario, 
it would be difficult to isolate the effect of cell-cell adhesion on proliferation. 
%we postulate that at intermediate E-cadherin expression levels there would be little to no correlation with patient survival. 
The prediction from our model in terms of tumor proliferation behavior is borne out 
by the limited analysis we have carried out i.e. role of E-cadherin as a tumor promoter or suppressor 
depends on the level of gene expression. %(see Fig.~\ref{figsi6}d). 

%\section*{IV. The spatial distribution of cell proliferation rate}
%We calculate the average cell proliferation rate, $\Gamma(r)$, at distance $r$ from the tumor center,
%\begin{equation}
%\Gamma(r) = \frac{N(r,t)-N(r,t-\delta t)}{\delta t},
%\end{equation}
%where $N(r,t)$ is the number of cells at time $t=650,000s$ and $\delta t=5000s$ is the time interval. 
%The average is over polar and azimuthal angles, for all cells between $r$ to $r+\delta r$. 
%%$\Gamma(r)$ is calculated at $t=650,000s$ and $\delta t=5000s$.  
%Closer to the tumor center, at low $r$, $\Gamma (r)\sim 0$ indicating no proliferative activity. 
%However, for larger values of $r$, proliferation rate rapidly increases approaching the periphery. 
%The proliferation rate is highest for $f^{ad}=1.5\times 10^{-4}$. 

\section*{IV. Simulation details}
At time, $t=0$, we begin with seeding 100 cells. The radii of these initial cells are distributed according to a Gaussian, 
$p(R_{i})=\frac{1}{0.5\sqrt{2\pi}} e^{-(R_{i}-4.5)^2/2\times 0.5^2}$. Similarly, the elastic moduli, $E_{i}$ and the Poisson ratio $\nu_{i}$ are also 
characterized by a Gaussian distribution with standard deviation of $10^{-4}\mathrm{MPa}$ and $0.02$ respectively (mean values are
given in Table S1).   
The receptor and ligand concentration on the cell surface are distributed according to a Gaussian 
($p(c_{i}^{rec}(c_{i}^{lig}))= \frac{1}{0.02\sqrt{2\pi}} e^{-(c_{i}^{rec}(c_{i}^{lig})-0.9)^2/2\times 0.02^2}$), centered around the mean (=0.9) with a dispersion of $0.02$. 
For subsequent cell division cycles, the newborn cell properties are sampled from the same distribution as detailed above. 
At each time step, the growth rate of the cell is also picked from a Gaussian distribution.  

The volume of growing cells increases at a constant rate $r_V$. 
Cell radii are updated from a Gaussian distribution with the mean rate $\dot{R} = (4\pi R^2)^{-1} r_V$ and dispersion of $10^{-5}$.  
Over the cell cycle time $\tau$, 
\begin{equation}
r_V = \frac{2\pi (R_{m})^3}{3\tau},
\end{equation}
where $R_{m}$ is the mitotic radius. See Ref.~\cite{malmi2018cell} for additional details.

\newpage

\newpage

\newpage

\newpage

\clearpage
\newpage

\begin{table}\centering
\caption{The values of the parameters used in the simulations. }

\begin{tabular}{ |p{7cm}||p{4cm}|p{5cm}|p{3cm}|  }
\hline
 \bf{Parameters} & \bf{Values} & \bf{References} \\
 \hline
Critical Radius for Division ($R_{m}$) &  5 $\mathrm{\mu m}$ & ~\cite{schaller2005multicellular}\\
 \hline
 Extracellular Matrix (ECM) Viscosity ($\eta$) & 0.005 $\mathrm{kg/ (\mu m~s)}$   & ~\cite{galle2005modeling}  \\
 \hline
 Benchmark Cell Cycle Time ($\tau_{min}$)  & 54000 $\mathrm{s}$  & ~\cite{freyer, casciari, landry}\\
 \hline
 Adhesive Strength ($f^{ad})$&  $0-3\times 10^{-4} \mathrm{\mu N/\mu m^{2}}$  & ~\cite{schaller2005multicellular}, This paper \\
 \hline
Mean Cell Elastic Modulus ($E_{i}) $ & $10^{-3} \mathrm{MPa}$  & ~\cite{galle2005modeling}    \\
 \hline
Mean Cell Poisson Ratio ($\nu_{i}$) & 0.5 & ~\cite{schaller2005multicellular}  \\
 \hline
 Death Rate ($b$) & $10^{-6} \mathrm{s^{-1}}$ & This paper \\
 \hline
Mean Receptor Concentration ($c^{rec}$) & 0.90 (Normalized) & ~\cite{schaller2005multicellular} \\
\hline
Mean Ligand Concentration ($c^{lig}$) & 0.90 (Normalized) & ~\cite{schaller2005multicellular}  \\
\hline
Adhesive Friction $\gamma^{max}$ &  $10^{-4} \mathrm{kg/ (\mu m^{2}~s)}$  &  This paper\\
\hline
Threshold Pressue ($p_c$) & $10^{-4} \mathrm{MPa}$  & ~\cite{schaller2005multicellular,montel2011stress}    \\
\hline
\end{tabular}
\end{table}

\clearpage
\newpage
\par
{\bf Movies:} In order to visualize the dynamic behavior of pressure experienced by cells, we generated movies from the simulations. 
All the movies show the three dimensional growth of cell collectives over $\approx$ 7.5 days. 
Indicated in the color scale with low pressures (blue) and high pressures (red) is the pressure experienced by a cell in units of MPa.
They demonstrate vividly the intercellular pressure fluctuations as the cells collectively expand. %We have quantified using various measures in the main text. 
Each movie frame is spaced at $1000$ seconds. The movies have been sped up by a factor of $2.16\times 10^{4}$, 
to aid visualization. The cell cycle time $\tau=\tau_{min}$.  
%Simulation movies can be found here \url{https://utexas.box.com/s/voitlbt4zv7l8ygc1g0n4ya5ob98octf}
Pressure relaxation of cells can be observed as the flickering of colors, indicating cells relaxing from high to low pressure or vice versa. 
Simulation movies can be found here \url{https://utexas.box.com/s/sl43kcoptciht61y4fm6e4klkagvtv5f} 
\par
{\bf Movie S1:} Intercellular pressure behavior at ${\bf f^{ad}=0}$: Colormap indicates the intercellular pressure. 
%Cells experiencing low pressure are in blue and those that experience high pressure are shaded red 
%(color bar in the video shows pressure in MPa). 
Cell division and death events are explicitly depicted. 
Supplementary Movie S1A shows the cross section through the cell collective at $f^{ad}=0$. 
Spatial pressure distribution shows elevated pressure in the interior which decreases towards the periphery.
\par
{\bf Movie S2:}  Pressure fluctuations in a growing cell collective at intermediate cell-cell adhesion strength ${\bf f^{ad} = 1.5\times 10^{-4} \mathrm{\mu N/\mu m^{2}}}$: 
Merging of two cell spheroids into a larger one can be observed. Growth of single cells and division events are depicted. 
Low pressure neighborhoods (depicted by blue color) is distributed throughout the surface of the spheroid. 
%Color bar shows cell intercellular pressure in MPa.  
Supplementary Movie S2A shows the cross section view at intermediate $f^{ad}$.
\par
{\bf Movie S3:} Intercellular pressure behavior at high cell-cell adhesion, ${\bf f^{ad}=3\times 10^{-4} \mathrm{\mu N/\mu m^{2}}}$: 
Pressure relaxation behavior for a growing cell collective at high cell-cell adhesion is visualized. 
Birth, apoptosis, growth and movement of cells is readily observed. 
Due to high cell-cell adhesion, groups of cells tend to form tightly packed clusters. Merging of such clusters 
can be seen.   
Supplementary Movie S3A shows the cross section view at high $f^{ad}$.
Pressure experienced by the cells decay as the tumor periphery is approached.

%\dataset{dataset_one.txt}{Type or paste caption here.}
%
%\dataset{dataset_two.txt}{Type or paste caption here. Adding longer text to show what happens, to decide on alignment and/or indentations for multi-line or paragraph captions.}

\bibliography{suppl}
\bibliographystyle{unsrt}